\documentclass[useAMS,usenatbib,usegraphicx]{mn2e}
\usepackage{txfonts}
\makeatletter
\DeclareMathAlphabet{\mathpzc}{OT1}{pzc}{m}{it}

\newcommand\h{\ensuremath{^\rmn h}}
\newcommand\m{\ensuremath{^\rmn m}}
\newcommand\tfwhm{\ensuremath{t_{1/2}}}
\newcommand\tE{\ensuremath{t_{\rm E}}}

\newcommand\rs{{\rm s}}
\newcommand\rb{{\rm b}}
\newcommand\final{3~}
\newcommand\finals{3~}
\newcommand\finalt{16~}
\newcommand\total{44554~}

\input epsf

\title[An Unrestricted Search for Microlensing Events towards M31.] {The
POINT-AGAPE survey II: An Unrestricted Search for Microlensing Events
towards M31}

\author[V.~Belokurov et al.]
{V. Belokurov,$^1$ J.~An,$^1$
N.~W.~Evans,$^1$
P.~Hewett,$^1$
P.~Baillon,$^2$
S.~Calchi~Novati,$^3$\newauthor
B.~J.~Carr,$^4$
M.~Cr\'ez\'e,$^{5,6}$
Y.~Giraud-H\'eraud,$^6$
A.~Gould,$^7$
Ph.~Jetzer,$^3$
J.~Kaplan,$^6$\newauthor
E.~Kerins,$^8$
S.~Paulin-Henriksson,$^6$
S.~J.~Smartt,$^{1,9}$
C.~S.~Stalin,$^6$
Y.~Tsapras$^4$ and M.~J.~Weston$^4$\newauthor
(The POINT-AGAPE\thanks{Pixel-lensing Observations on the Isaac
Newton Telescope - Andromeda Galaxy Amplified Pixels Experiment}
collaboration)\\
$^1$Institute of Astronomy, University of Cambridge,
Madingley Road, Cambridge CB3 0HA, UK\\
$^2$European Organization for Nuclear Research CERN,
CH-1211 Gen\`eve 23, Switzerland\\
$^3$Institut f\"ur Theoretische Physik, Universit\"at Z\"urich,
Winterthurerstrasse 190, CH-8057 Z\"urich, Switzerland\\
$^4$Astronomy Unit, School of Mathematical Sciences, Queen Mary,
University of London, Mile End Road, London E1 4NS, UK\\
$^5$Universit\'e Bretagne-Sud, Campus de Tohannic, BP 573,
F-56017 Vannes Cedex, France\\
$^6$Laboratoire de Physique Corpusculaire et Cosmologie, Coll\`ege de
France, 11 Place Marcelin Berthelot, F-75231 Paris, France\\
$^7$Department of Astronomy, Ohio State University,
140 West 18th Avenue, Columbus, OH 43210, USA\\
$^8$Astrophysics Research Institute, Liverpool John Moores University,
12 Quays House, Egerton Wharf, Birkenhead CH41 1LD, UK \\
$^9$Department of Pure and Applied Physics, The Queen's University of
Belfast, Belfast, BT7 1NN, Northern Ireland}

\begin{document} 

\maketitle
\begin{abstract}
An automated search is carried out for microlensing events using a
catalogue of \total variable superpixel lightcurves derived from our
three-year monitoring program of M31. Each step of our candidate
selection is objective and reproducible by a computer. Our search
is unrestricted, in the sense that it has no explicit timescale
cut. So, it must overcome the awkward problem of distinguishing
long-timescale microlensing events from long-period stellar variables.

The basis of the selection algorithm is the fitting of the superpixel
lightcurves to two different theoretical models, using variable star
and blended microlensing templates.  Only if microlensing is preferred
is an event retained as a possible candidate. Further cuts are made
with regard to (i) sampling, (ii) goodness of fit of the peak to a
Paczy\'nski curve, (iii) consistency of the microlensing hypothesis
with the absence of a resolved source, (iv) achromaticity, (v)
position in the colour-magnitude diagram and (vi) signal-to-noise
ratio. Our results are reported in terms of first-level candidates,
which are the most trustworthy, and second-level candidates, which are
possible microlensing but have lower signal-to-noise and are more
questionable. The pipeline leaves just \final first-level candidates,
all of which have very short full-width half-maximum timescale
($\tfwhm <5$ days) and \finals second-level candidates, which have
timescales $\tfwhm = 31, 36$ and 51 days.  We also show \finalt
third-level lightcurves, as an illustration of the events that just
fail the threshold for designation as microlensing candidates. They
are almost certainly mainly variable stars.

Two of the \final first-level candidates correspond to known events
(PA 00-S3 and PA 00-S4) already reported by the POINT-AGAPE project.
The remaining first-level candidate is new. This algorithm does not
find short-timescale events that are contaminated with flux from
nearby variable stars (such as PA 99-N1).
\end{abstract}

\section{Introduction}

Microlensing experiments towards M31 have now been in progress for a
number of years and are beginning to report results. Ongoing surveys
include the POINT-AGAPE (e.g., Auri\`ere et al. 2001;
Paulin-Henriksson et al. 2003; Calchi Novati et al. 2003), the WeCAPP
(Riffeser et al. 2001, 2003) and the MEGA (de Jong et al. 2004)
collaborations.

The dataset used by the POINT-AGAPE collaboration consists of three
seasons (1999-2001) of imaging taken with the Wide Field Camera on the
{\it Isaac Newton Telescope} (INT). In each season, data were taken in
two passbands ($r$, together with one of $g$ and $i$) for one hour per
night for roughly sixty nights during the six months that M31 is
visible at low airmass. The field of view of the Wide Field Camera is
33\arcmin$\times$33\arcmin\ . There are two fields, located north and
south of the centre of M31, as shown in Figure 1 of An et
al. (2004b). The field locations are motivated by the suggestion of
Crotts (1992) that the event rate to sources in the near and far disks
is different. The lines of sight to the far disk as compared to the
near disk are longer and pass through more of the dark halo. So, if
the halo dark matter is in compact form, this should manifest itself
in a greater microlensing rate towards sources in the far disk.  We
have already found a number of interesting individual microlensing
events towards M31 (e.g., Auri\`ere et al. 2001; Paulin-Henriksson et
al. 2002, 2003; An et al. 2004a), carried out a survey for classical
novae (Darnley et al. 2004) and reported the locations, periods and
brightness of $\sim 35000$ variable stars (An et al. 2004b). The
variable stars also show a substantial asymmetry between the near and
far disk caused by the effects of extinction and dust. So, An et
al. (2004b) proposed a new idea -- namely, that it is better to look
for an east-west asymmetry in microlensing events rather than a
near-far asymmetry. The variable star distributions are east-west
symmetric to a good approximation.

In the context of the POINT-AGAPE experiment, there are three relevant
goals for a survey of microlensing events. First, one might want to
identify any genuine microlensing events (without regard to the
selection criteria). This is a worthy goal, as the microlensing events
are of interest in their own right (e.g., Auri\`ere et al. 2001; An et
al.  2004a).  Second, one might want to identify candidate
microlensing events such that the detection probability is spatially
homogeneous. This would enable any asymmetry present to be empirically
constrained. Third, one might want to identify candidate microlensing
events such that the probability that a specified microlensing event
is detected can ultimately be computed. Achieving the third aim
involves substantially more effort than the first aim. It may also
carry the penalty that some genuine microlensing events are omitted
from the final list of candidates.

Although most simple simulations of unresolved M31 microlensing
experiments predict that the typical full-width half-maximum timescale
($\tfwhm$) is quite short, it is desirable not to restrict ourselves
to select short-timescale candidates a priori. However, the optimum
choice of candidate selection algorithm may be different, according to
whether the sought-after events have long or short timescale. A
serious difficulty in identifying long-timescale events is
contamination of the sample from long-period variable stars (e.g.,
Miras), which -- given the sampling -- may not repeat during the 3
year baseline of the experiment. For short-timescale events, confusion
with stellar variables is less problematic. Short-period variables
(e.g., Cepheids) show multiple bumps in their lightcurves over the
three-year baseline. For this reason, it makes sense to devise
different algorithms for candidate selection for long and short
timescale events.

Paulin-Henriksson et al. (2003) have reported 4 candidates in a survey
for high signal-to-noise ratio, short-timescale microlensing events in
the first two years of INT data. Specifically, they restricted
attention to events with a full-width half-maximum timescale $\tfwhm <
25$ days and with a flux variation $\Delta f$ in the $R$ band that
exceeds that of an $R= 21$ magnitude star (i.e., $R(\Delta f) <
21$). More recently, Paulin-Henriksson et al. (2004) and Calchi Novati
et al. (2004) have extended their search to the full 3 year dataset.
The advantage of the timescale and magnitude cuts is that the
microlensing nature of bright, short events is usually unambiguous.
The disadvantage is that the cuts restrict attention to $\sim 15 \%$
of all the variable lightcurves in the POINT-AGAPE database.

The purpose of this paper is to complement the short-timescale
searches of Paulin-Henriksson et al. (2003, 2004) and Calchi Novati et
al. (2004) by devising a candidate selection algorithm that is not
limited to short-timescale events. We remove all restrictions on
full-width half-maximum timescale and search for fainter flux
variations in the entire catalogue ($19 < R(\Delta f) < 24$). We
pay particular attention to the difficulty of distinguishing
long-timescale microlensing events from long-period Miras. The
selection algorithms of Paulin-Henriksson et al. (2003, 2004) and this
paper are entirely independent, although both the imaging data and the
constructed lightcurves are the same.  In both Paulin-Henriksson et
al. (2003, 2004) and this paper, the aim is to carry out a fully
automated survey for which the efficiency can ultimately be
calculated.

This paper begins with a description of the catalogue of superpixel
lightcurves constructed from our monitoring program of M31 during
1999-2001. Then, the selection procedure for microlensing candidates
is described in \S 3. The M31 source stars are generally not resolved
and so the selection algorithm is different from those employed in
classical microlensing experiments such as MACHO (Alcock et al. 1997)
and EROS (Aubourg et al. 1995). The survivors of the seven cuts of the
selection procedure are a set of \final first-level candidates and
\finals second-level candidates.  The first-level candidates are the
most trustworthy and, notably, are all short.  Their lightcurves,
locations and microlensing parameters are presented in \S 4.  The
\finals second-level candidates form a more loose selection of
possible, if questionable, events.  Also extracted are a set of
\finalt third-level lightcurves that lie just below our threshold for
designation as microlensing candidates.  They are discussed in \S 5.

\begin{table}
\begin{center}
\begin{tabular}{lcclcc}
\hline
Field & CCD & Fraction & Field & CCD & Fraction \\
\hline
northern & 1  & 0.0987 & southern & 1 & 0.0611 \\ 
northern & 2  & 0.0766  & southern & 2 & 0.0772 \\
northern & 3 & 0.0438 & southern & 3 & 0.1582 \\
northern & 4 & 0.0702 & southern & 4 & 0.0598 \\
\hline
\end{tabular}
\end{center}
\caption{The fractions of each field and CCD removed
 by the masking of
the resolved stars on the reference frame. The field and CCD placings
are shown in Figure 1 of An et al. (2004b).}
\label{tab:area}
\end{table}

\section{The Construction of the Catalogue}

The $r$ band lightcurves are available for all three seasons. The $g$
band monitoring was largely discontinued after the first season (1999)
and replaced with $i$ band monitoring during 2000 and 2001. There are
some, albeit few, $i$ band data for 1999. Full details of the data
acquisition and analysis can be found in An et al. (2004b). As in that
paper and in Paulin-Henriksson et al. (2003, 2004), the superpixel
method is used to process the data and construct the lightcurves. The
basic idea is to smooth the data with a median filter to estimate the
background, to apply an empirical correction for seeing variations and
hence to build lightcurves for $7 \times 7$ superpixels. Henceforth,
we deal exclusively with the set of derived superpixel lightcurves
rather than the images themselves.  We select the variable superpixel
lightcurves by requiring that they have at least 3 consecutive
datapoints at least 3$\sigma$ above the baseline in the $r$ band. This
gives us a raw catalogue of 97280 $r$ band selected variable
lightcurves.

The typical seeing (full-width half-maximum) for the INT data exceeds
3 pixels and the range of seeing present in even the processed data is
a full factor of two. The superpixel method involves a block-averaging
scheme with a scale-length of 7 pixels ($2\farcs3$).  One of the
drawbacks of the method is that it does not work when a pixel is
within approximately the smoothing length of either bad pixels or 
bright resolved stars. Therefore, the superpixel lightcurves within
$\sim 6$ pixels of resolved objects are full of "variations" due to
the changing seeing or variability of the nearby resolved object.
Thus, the statistical properties of the lightcurves from within this
portion of the CCD are very different from those that are
well-separated from resolved objects.

An et al. (2004b) removed such spurious lightcurves by utilising
spectral analysis based on Lomb's periodogram.  Here we adopt a
different procedure. This is partly because we need to calculate the
actual area of M31 covered by our survey for the purposes of
estimating the efficiency, and partly because short $\tfwhm$
microlensing candidates fail to pass the spectral analysis cut of An
et al. (2004b).  We construct a mask of the known CCD defects,
together with regions around all resolved stars detected in the
reference frame. We note that masking the bad pixels is intrinsically
different from masking the areas around the resolved stars, as the
former is fixed with respect to the CCD, while the latter is fixed with
respect to the sky. Whilst the stability of the pointing of the INT is
good, the pointing nonetheless does vary slightly from frame to
frame. The masking is therefore applied in two stages.

First, the areas within 6 pixels of any star brighter than $R=20.5$,
together with annuli between 1.5 pixels and 3 pixels around stars
fainter than $R= 20.5$ are incorporated into the mask. We chose these
numbers after examining Figure 7 of An et al. (2004b), which shows a
surfeit of variable lightcurves at these pixel distances from resolved
stars. Table~\ref{tab:area} shows the fraction of area lost by masking
for each field and CCD combination.  After masking, \total variable
superpixel lightcurves remain. Over half the original lightcurves have
been removed, though they occur on $< 10$ \% of the total surveyed
area. This illustrates the fact that the statistical properties of the
removed lightcurves are very different from those that remain. This is
why it is beneficial to have excised them from the catalogue, even
though -- as we will see -- some bona fide microlensing events have
been lost through masking. In principle, this difficulty can be
significantly reduced by techniques such as difference image analysis,
albeit at some computational cost.

Second, the registration of each frame is known, so the CCD
coordinates corresponding to any superpixel at that epoch can be
calculated. If these lie on a known CCD defect, the datapoint is
eliminated from the lightcurve.  The final catalogue contains \total
cleaned variable lightcurves, in which all datapoints lying on bad
pixels have been removed. This is the catalogue through which we
search for microlensing events.

\begin{figure*}
\vspace{0.24\hsize}\centering{\tt fig1.gif}\vspace{0.25\hsize}
\caption{Each lightcurve is plotted in the ($\log \Delta \chi^2_{\rm
micro}, \log \Delta \chi^2_{\rm var}$) space. Notice the dense black
cloud representing variable stars. The first cut (red line) is given
by equation~(\ref{eq:firstcut}) and is chosen to excise this cloud.
Lightcurves passing the first cut are plotted as red dots. For
comparison, the statistical $F=1$ test of
equation~(\ref{eq:comparison}) is shown as a blue line, while the
$F=50$ test is shown as a green line.}
\label{fig:chisq}
\end{figure*}
\begin{figure*}
\includegraphics[width=12cm]{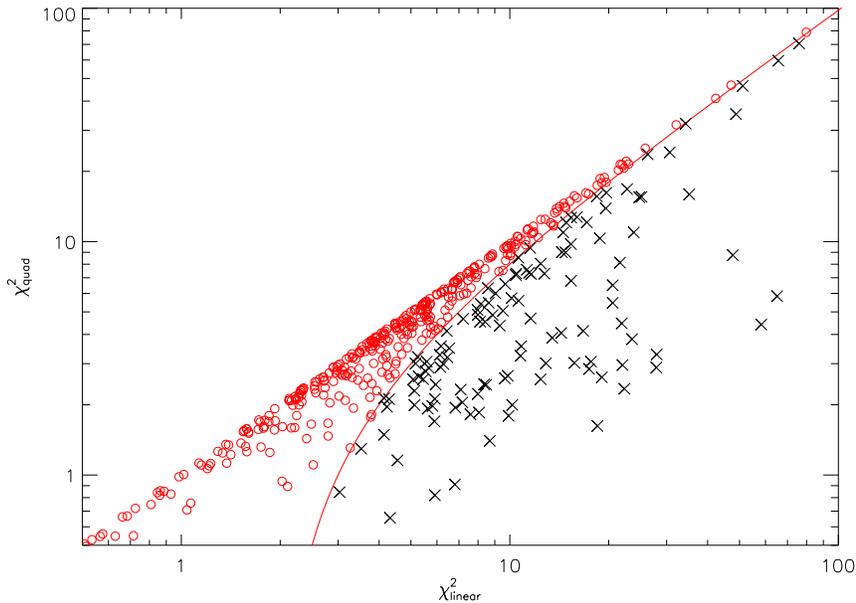}
\caption{The 488 lightcurves that survive after the fourth cut are
then subjected to an achromaticity test. The flux change in two
passbands is fitted by a linear and quadratic law and the lightcurves
are plotted in the ($\chi^2_{\rm linear}, \chi^2_{\rm quad}$)
space. The red circles represent the lightcurves that satisfy the
achromaticity cut given in equation~(\ref{eq:achrom}).}
\label{fig:chromcut}
\end{figure*}
\begin{figure*}
\vspace{0.24\hsize}\centering{\tt fig3.gif}\vspace{0.25\hsize}
\caption{Variable stars are represented as grey dots on the analogue
of the colour-magnitude diagram. The 369 survivors after the fifth cut
are plotted as red dots.  The sixth cut (dotted blue line) given in
equation~(\ref{eq:cmdcuts}) is a colour cut that removes the red-most
portion of the colour-magnitude diagram dominated by Miras. Note that
$R(\Delta f) - I(\Delta f)$ and $I(\Delta f)$ are determined from the
data using the transformations in Appendix A.}
\label{fig:cmds}
\end{figure*}
\begin{figure*}
\includegraphics[width=12cm]{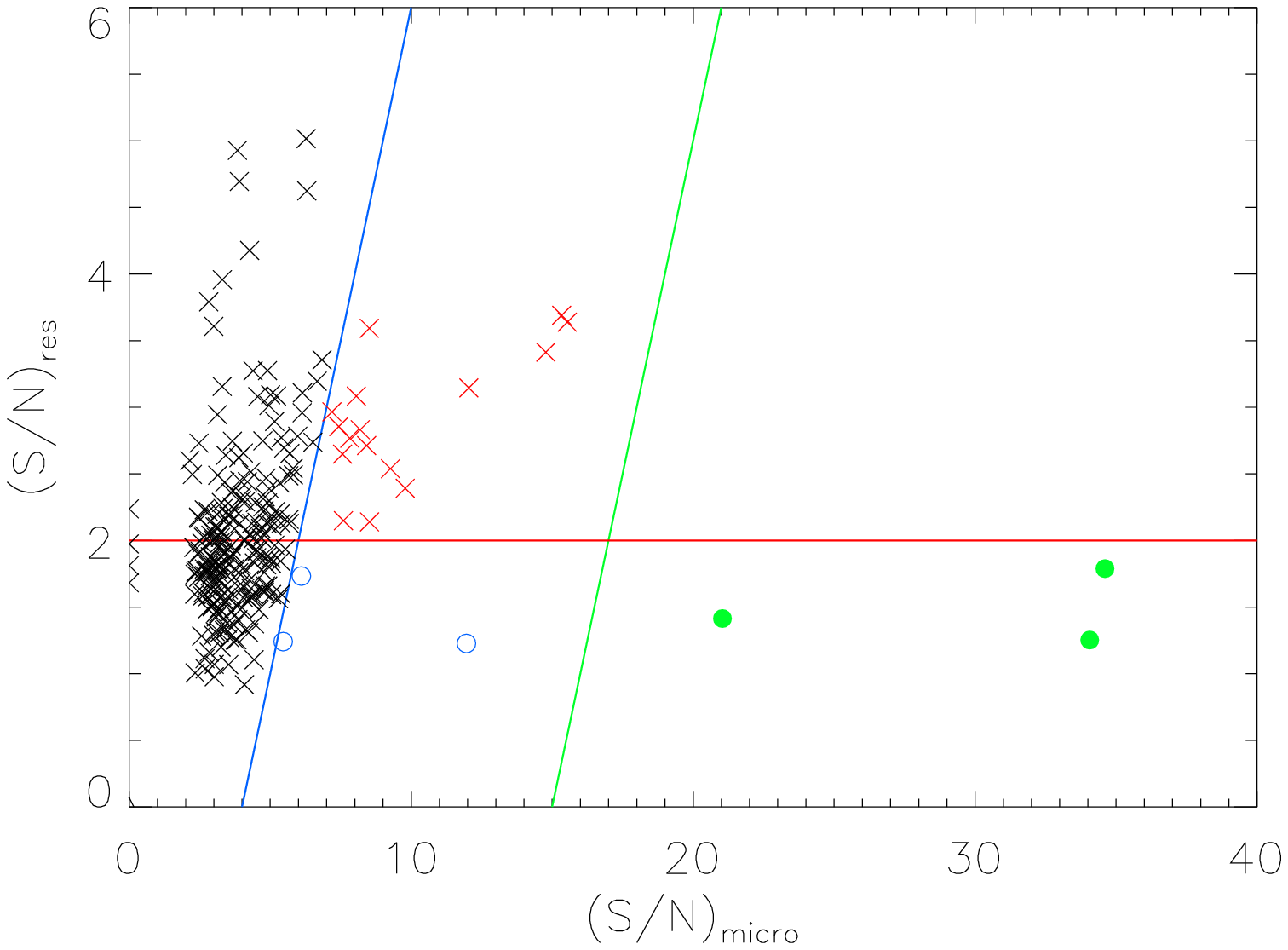}
\caption{The 273 lightcurves that survive after the sixth cut are then
plotted in the $(S/N)_{\rm res}, (S/N)_{\rm micro}$ plane.  There is a
tail of events with microlensing signal-to-noise ratio dominating over
the signal-to-noise ratio in the residuals. A set of \final
first-level candidates (filled green circles) are extracted by the
stringent cut given in equation~(\ref{eq:stona}). Also shown are a
looser set of a \finals second-level candidates (open blue circles),
which are extracted by the weaker cuts given in
equation~(\ref{eq:stonb}). Finally, a set of \finalt third-level
lightcurves (red crosses) are extracted by the cuts given in
equation~(\ref{eq:stonc}).}
\label{fig:ston}
\end{figure*}

\section{The Microlensing Selection Criteria} 

\subsection{Overall Strategy}

The selection criteria used to identify microlensing lightcurves in
experiments such as MACHO require just one significant bump in the
lightcurve, generally by insisting on goodness of fit to the standard
Paczy\'nski curve for a point source microlensed by a point lens
(e.g., Alcock et al. 1997). In the POINT-AGAPE experiment, pixel
lightcurves may receive contributions from more than one variable
source located in the same superpixel, and so the variety of
lightcurves is very rich. The development of selection criteria for
identifying microlensing events from a dataset of pixel lightcurves
involves different challenges from those presented by conventional
microlensing experiments in which the sources are resolved.

In this paper, the focus is on carrying out a search that has no
explicit restrictions on timescale. This necessitates the devising of
algorithms to distinguish long-timescale microlensing events from
contaminating long-period variable stars.  Goodness of fit to standard
microlensing lightcurves alone cannot resolve this problem.  The
selection algorithm begins by comparing lightcurves to two different
theoretical models -- namely, a simple variable star template and a
standard Paczy\'nski microlensing curve. Only if the microlensing fit
is better than the alternative is a lightcurve retained as a possible
candidate. Although we do not pursue it here, this line of reasoning
is capable of further elaboration; for example, a very general
algorithm for identifying microlensing events in pixellated data might
begin by comparing goodness of fit to a number of competing models
(e.g., variable stars, novae, supernovae, microlensing, microlensing
with a bumpy baseline caused by variable contaminants).

Cuts based on goodness of fit and absence of multiple bumps are never
enough to eliminate variable stars.  Worryingly for the subject of
microlensing, there are lightcurves that are extremely well-fit by
blended microlensing curves in two or more pass-bands, but that are
almost certainly Miras or other long-period variables. Achromaticity
tests are a conventional way of distinguishing stellar variability
from microlensing.  We find that achromaticity tests with the $r$ and
$i$ band data are effective, but tests with the $r$ and $g$ band data
are less powerful, mainly because most variables are red stars and so
have low $g$ band flux. Cuts in an analogue of the colour-magnitude
diagram can also be useful in excising contaminants from our dataset,
although care is needed to avoid eliminating too many possible
sources.

It is also important that every stage in the selection of microlensing
candidates is objective and reproducible by a computer, otherwise the
efficiency of the process cannot be calculated by Monte Carlo
simulations. The only permissible application of cuts by eye is to
eliminate phenomena that are not included in the simulations -- such
as lightcurves associated with cosmic rays, satellite trails or CCD
defects. In our work, any such lightcurves are eliminated (sometimes
by visual examination of the original images) before the construction
of the input catalogue of \total lightcurves.  We remark that
the M31 microlensing survey carried out by de Jong et al. (2004)
involved a final cull by eye at the end of candidate selection
(reducing the number of candidates from 126 to 12). This is not
reproducible by computer and so the efficiency of the de Jong et al.
survey cannot be determined by a straightforward computer algorithm.

\begin{table*}
\begin{center}
\begin{tabular}{lcccccccccc}
\hline
Cut & Description &F1, CCD1 & F1, CCD2 & F1, CCD3 & F1, CCD4 
& F2, CCD1 & F2, CCD2 & F2, CCD3 & F2, CCD4 & Total \\
\hline
& &	 6474&	6518&	4204&	5498&	2620&	4705&	8971&   5564&	44554\\
1 &Global Fit&	 449&	609&	863&	685&	213&	425&	948&	719&	4911\\
2 &Sampling&	 175&	257&	275&	222&	71&	201&	311&	206&	1718\\
3 &Local Fit&	 96&	82&	50&	84&	29&	85&	124&	62&	614\\
4 &Unresolved Source &	 77&	60&	35&	64&	23&	70&	105&	54&   488\\
5 &Achromaticity &	 65&	49&	25&	45&	16&	50&	75&	44&   369\\
6 &Colour-Magnitude&       44&    33&     21&     34&      12&     32&
61&    36&   273\\ \hline
7 &Signal-to-Noise: First Level&       0&     0&      0&      0&       0&       1&      1&      1&   3\\
7 &Signal-to-Noise: Second Level&       0&     0&      0&      0&       1&       0&      2&      0&   3\\
7 &Signal-to-Noise: Third Level&       2&     2&      0&      4&       0&       2&      3&      3&   16\\
\hline
\end{tabular}
\end{center}
\caption{The numbers of lightcurves surviving after each cut, broken
down according to their occurrence in the CCDs and fields. F1 denotes
the northern field, F2 the southern field.  The original catalogue
contains \total variable lightcurves before the first cut is applied.
The seventh cut is based on signal-to-noise ratio in microlensing
versus residuals and is used to extract first-level, second-level and
third-level lightcurves. The first-level and second-level lightcurves
are microlensing candidates, the third-level are probably almost all
variable stars.}
\label{tab:events}
\end{table*}

\subsection{An Automated Algorithm for Candidate Selection}

All lightcurves are automatically fitted with two models; (1) a
Paczy\'nski curve with a flat baseline and (2) a crude variable star
lightcurve, using all the datapoints in all three passbands. The first
fit has nine parameters (namely $f_{\rs,g}$, $f_{\rs,r}$, $f_{\rs,i}$,
the source fluxes in $g, r$ and $i$; $f_{\rb,g}$, $f_{\rb,r}$,
$f_{\rb,i}$, the blending fluxes in $g, r$ and $i$; $u_0$ the impact
parameter; $t_{\rm max}$ the time of maximum and $\tE$ the Einstein
radius crossing time). The fitted parameters are known to be
degenerate in blended events (e.g., Wo\'zniak \& Paczy\'nski 1997),
but this is not important at such an early stage in the selection
algorithm. There are some restrictions on the values of the fitted
parameters -- namely, all the fluxes must be positive, the impact
parameter is restricted to lie between 0 and 2, and the Einstein
radius crossing time $\tE$ must lie between 1.5 days and 1.5 yrs.  We
refer to this fit as the global microlensing fit.  The second model to
be fitted is a single sinusoid. Of course, most variable stars have
more complicated lightcurves than sinusoids, but this crude model is
adequate for our purpose here. The fit has eight parameters (the
baseline and the amplitude in $g$, $r$, and $i$, the phase, the
period). However, in practice, we do not fit the period as a freely
selectable parameter, but fix it to one of thirty values and fit the
rest of parameters. The trial set of periods is determined as follows.
The data are first ``robustified'' (Alcock et al. 2000) by removing
the five highest and lowest values.  The robustified data for each
passband are analyzed with Lomb's periodogram (see e.g., Press et
al. 1992) and the values of the frequency corresponding to the highest
10 peaks in each of the 3 passbands are extracted. As before, all
fluxes are restricted to be positive in the fit. We refer to this fit
as the variable star fit.

From these two fits, we can compute the improvement of $\chi^2$ for
the Paczy\'nski curve fit over a flat baseline ($\Delta \chi^2_{\rm
micro} = \chi^2_{\rm bl} - \chi^2_{\rm micro}$) and the similar
$\chi^2$ improvement of the variable star lightcurve fit over a flat
baseline ($\Delta \chi^2_{\rm var} = \chi^2_{\rm bl} - \chi^2_{\rm
var} $). In other words, both $\Delta \chi^2_{\rm micro}$ and $\Delta
\chi^2_{\rm var}$ are the $\chi^2$ improvement (the absolute value of
the decrease in $\chi^2$) so that a larger number means a better fit.
As shown in Figure~\ref{fig:chisq}, there is an enormous cloud in the
($\Delta \chi^2_{\rm micro}, \Delta \chi^2_{\rm var}$) plane
corresponding to variable stars. The {\bf first cut} is chosen so as
to excise this cloud by insisting that
\begin{equation}
\Delta \chi^2_{\rm var} < 0.75 \Delta \chi^2_{\rm micro}.
\label{eq:firstcut}
\end{equation}
This empirically derived cut is shown as the red line in
Figure~\ref{fig:chisq}. It ruthlessly reduces the candidate
lightcurves by an order of magnitude from \total to 4911. 

It is instructive to compare this approach with the classical
statistical method used for hypothesis testing, namely the $F$ test
(e.g., Lupton 1993, chap. 12). Suppose we wish to distinguish between
two models (here variable stars and microlensing) that are both fit to
the same number of datapoints, but have different numbers of free
parameters. Then, we construct $\Delta \chi^2 = \chi^2_{\rm var} -
\chi^2_{\rm micro} = \Delta \chi^2_{\rm micro} - \Delta \chi^2_{\rm
var}$ and compare this to the difference in the number of free
parameters $\Delta n$.  If the additional parameters are statistically
redundant, the $\chi^2$ difference $\Delta \chi^2$ distributes as a
$\chi^2$ distribution with $\Delta n$ degrees of freedom.  Then,
roughly speaking, provided that $\Delta \chi^2\gg\Delta n$, one can
conclude that the added free parameters are significant and thus that
the model with the greater number of parameters is favoured over the
one with fewer. The exact criterion, which can be formally written as
\begin{equation}
\Delta\chi^2> F (\Delta n),
\label{eq:comparison}
\end{equation}
is dependent on the chosen significance level. For the case in hand,
$\Delta n=1$. So, assuming the reported photometric errors are
well-determined and random, the 1-$\sigma$ (68\%) confidence
corresponds to $F=1$.  However, under a conservative approach, $F=50$
is probably more reasonable, given the characteristics of our data.
The cuts corresponding to $F=1$ and $F=50$ are shown as blue and green
lines in Figure~\ref{fig:chisq}. They run through the cluster of
variable stars, rather than just below it, so they are not optimal.
The cut that we actually apply (equation~\ref{eq:firstcut}) is both
more stringent and more efficient, as it moves the intercept downwards
to just below the cluster of variable stars.

The {\bf second cut} demands that there are datapoints on both the
rising and falling part of the lightcurve.  Specifically, we require
two datapoints within 1.5 \tfwhm\ of the peak on each side, and we
require more than 5 datapoints within 6 \tfwhm\ in one passband, plus
at least 1 further datapoint in a second passband.  All these
datapoints must be at least $3\sigma$ away from the baseline as judged
by the global microlensing fit. This may lose some bona fide
microlensing events, for which incomplete coverage means that we
cannot be certain of the nature of the event.

The {\bf third cut} proceeds by carrying out a local blended
Paczy\'nski fit to all datapoints within 6 \tfwhm. This is done in as
many passbands as possible, usually two as the data are not generally
available in all three bands during the event.  The rationale for a
local Paczy\'nski fit is that that the global fit may not return the
proper baseline because of contamination by nearby variable
stars. However, even if the baseline is bumpy, we expect that the
maximum is a good fit to the theoretical curve. This is because the
typical timescale for the contaminating variable behaviour is much
longer than the duration of microlensing, so that the baseline can be
approximated as constant during the event.  The fluxes are again
restricted to be positive and the impact parameter must lie between 0
and 2. The Einstein radius crossing time $\tE$ may take any value
between 1.5 days and 1.5 years.  We refer to this fit as the local
microlensing fit. If the $\chi^2_{\rm local}$ per degree of freedom is
greater than 2, then the event is rejected.

\begin{figure}
\begin{center}
\includegraphics[height=10cm]{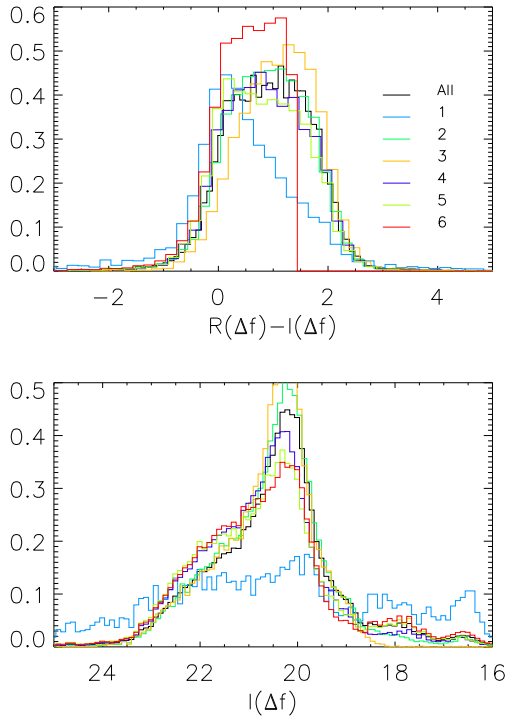}
\end{center}
\caption{For each cut applied singly to the \total lightcurves in the
catalogue, the distributions of the survivors are shown as a function
of Cousins colour $R(\Delta f) - I(\Delta f)$ and magnitude
$I(\Delta f)$. The cuts are:1 $\Delta \chi^2$ (in cyan), 2 sampling
(in green), 3 local fit (in brown), 4 resolved source (in blue),
5 achromaticity (in yellow) and 6 colour-magnitude (in red). The
distributions in the starting catalogue are shown in black. There is
no bias introduced via any of the cuts -- except of course the
colour-magnitude cut which introduces a cut-off at $R(\Delta f) -
I(\Delta f) = 1.5$.}
\label{fig:cuts}
\end{figure}

The {\bf fourth cut} follows from insisting that the global
microlensing fit is consistent with the fact that the source star must
be unresolved.  Recall that all the resolved stars were removed from
our starting catalogue by our mask. Therefore, none of the candidates
corresponds to a resolved star in the reference frame.  In particular,
global microlensing fits with low maximum magnification $A_{\rm max}$
may predict that the source is bright and resolved, which would make
the microlensing interpretation inconsistent. It is straightforward to
derive the detection limit for a resolved star, assuming (i) that it
is limited by the Poisson noise of the `sky', (ii) that the baseline
counts of the superpixels ($7\times7$ pixel$^2$) are dominated by the
same `sky' as that on the reference frame and (iii) that the point
spread function is an isotropic Gaussian.  The limit of the source
flux to be $n\sigma$ above the sky noise is
\begin{equation}
f_{\rm limit} = {n a \sqrt{\upi} 
\over 14(1-2^{-(a/s)^2})} \sqrt{{f_{\rm bl} \over E}},
\label{eq:jins}
\end{equation}
where $E$ is the exposure time, $f_{\rm bl}$ is the baseline of the
global microlensing fit, $s$ is the seeing disk (FWHM) in pixels and
$a$ is the aperture (diameter) size in pixels. The seeing and exposure
times of the reference frames are listed in Table 1 of An et
al. (2004b). Assuming that $a \approx s$ then gives us values for all
the physical quantities on the right-hand side of
equation~(\ref{eq:jins}).  Clearly, there is more noise than pure
photon noise, and thus, to be conservative, we choose n=10 for the
detection limit;
\begin{equation}
f_{\rm limit} = {10 s\sqrt{\upi} \over 7} \sqrt{{f_{\rm bl} \over E}}.
\label{eq:bloodylimit}
\end{equation}
Here, we convert the unit of fluxes into ADU s$^{-1}$ so that,
strictly speaking, the actual limit corresponds to $n$ times the
square root of the gain (in e ADU$^{-1}$).  The fourth cut is to
require that the 1$\sigma$ lower bound of the $r$ band source
flux~\footnote{The source flux found from the fit is actually the
fraction of the source light falling within the superpixel and hence
is a lower limit to the true source flux. By ignoring this correction
($\approx 15 \%$ for 1\farcs5 seeing), our cut is somewhat looser in
practice than reported in the text.}  of the global microlensing fit
be less than the flux limit of equation~(\ref{eq:bloodylimit}). There
are well-known degeneracies in the fitting of blended microlensing
events, with the consequence that the source flux from the fit is not
wholly reliable. This is why such a cautious choice is made in the
resolved flux limit in equation~(\ref{eq:bloodylimit}). The fitting
routine can mimic flatter-topped lightcurves of a population of
variable stars with low magnification microlensing lightcurves.  The
effect of this cut is to removes such spurious, large impact parameter
events.

The {\bf fifth cut} is an achromaticity test. As microlensing is
intrinsically achromatic, the ratio of the flux difference over the
baseline (that is, the variable flux) between two different passbands
must be constant over time (e.g., Ansari et al. 1997, Auri\`ere et
al. 2001). The same is not true for most pulsating stellar variables,
which usually become bluer as they approach maximum (e.g., Sterken \&
Jaschek 1996). The datapoints in two pass-bands along the bump are
binned, with the bin-size depending on the width of the bump and the
number of available datapoints. A straight line and a quadratic
polynomial are then fitted to the flux measurements in two passbands.
If a linear fit is preferred (i.e., the quadratic coefficient is
statistically redundant), then this suggests that the lightcurve is
achromatic.  The fifth cut is to insist that the linear fit is
preferred, namely
\begin{equation}
\Delta \chi^2 = \chi^2_{\rm linear} - \chi^2_{\rm quad} < 2.
\label{eq:achrom}
\end{equation}
For events that peak in the first season, the $r$ and $g$ band data
are used. For all other events, the $r$ and $i$ band data are used.
The effect of the achromaticity cut is shown in
Figure~\ref{fig:chromcut}.

\begin{figure*}
\includegraphics[width=6cm, angle=270]{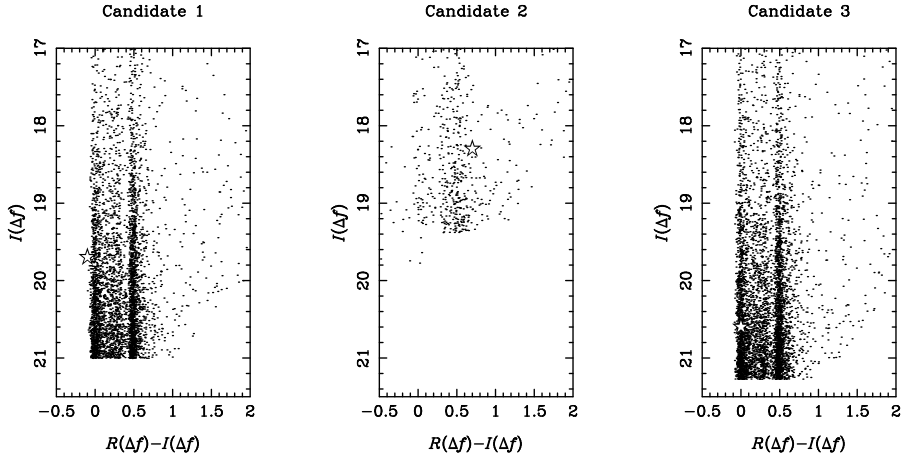}
%\vskip 10cm
\caption{Synthetic Cousins $I(\Delta f)$ versus $R(\Delta f)-I(\Delta
f)$ colour-magnitude diagrams showing the final \final first-level
candidates (star symbols) and their corresponding local source
populations. For candidates 1 and 3 we assume a disk population,
whilst for candidate 2 we assume the source belongs to the bulge. The
local source populations are generated by assuming a peak S/N
threshold of 20 (see main text).  The size of the star symbols 
reflect the $1\,\sigma$ error of $\sim 0.1$~mag in the transformed
photometry. (Note that the locations of the candidates are deduced
using the microlensing fit, and so differ slightly from the locations
in Fig.~3.)}
\label{fig:eamonn}
\end{figure*}

The {\bf sixth} cut is the elimination of the most troublesome
long-period variable stars that show only one bump in the POINT-AGAPE
dataset, using an analogue of the colour-magnitude diagram. As a
measure of the brightness of each variable, An et al. (2004b)
introduced the ``pseudo-magnitude''. Because the fraction of light
contributed by a variable object to the total superpixel flux at a
given epoch is not known, the real magnitude of any variable cannot be
determined. Under these circumstances, we measure the quantity $\Delta
f$, which is the flux variation between minimum and maximum in ADU
s$^{-1}$. We then convert $\Delta f$ into magnitudes using $r(\Delta
f)=r_0-2.5\log_{10}(\Delta f)$, where $r_0$ is the zeropoint, that is,
the $r$ band magnitude of a star whose flux is 1 ADU s$^{-1}$ in the
reference image.  This quantity $r(\Delta f)$ --- or $i(\Delta f)$,
$g(\Delta f)$ as appropriate --- is called the
pseudo-magnitude. Similarly, the difference between two
pseudo-magnitudes, such as $r(\Delta f) - i(\Delta f)$, is a proxy for
the colour.  This allows us to build an analogue of the classical
colour-magnitude diagram for pixellated data, as shown in
Figure~\ref{fig:cmds}.  Note that, unlike the similar diagram
presented in figure 15 of An et al. (2004b), which is based on the
instrumental Sloan system, here we have converted the Sloan magnitudes
to the Cousins magnitudes using the flux calibration equation derived
in Paulin-Henriksson (2002) and listed in Appendix A.  Also plotted on
Figure~\ref{fig:cmds} as red dots are the survivors after the fifth
cut.  Most of variable stars appearing in the diagram are long-period
($\ga 100\ \mbox{days}$) Mira and semi-regular variables, which are
also the most insidious contaminants for the purpose of identifying
microlensing events. We remove the red-most candidates which occupy
the region of the Mira variables by imposing the colour cut
\begin{equation}
R(\Delta f)-I(\Delta f)>1.5,
\label{eq:cmdcuts}
\end{equation}
which is also shown as a straight line on Figure~\ref{fig:cmds}. Note
that this cut is not applied if there are only $r$ and $g$ band
data. The Miras do not form a sufficiently distinct population in the
$r(\Delta f)$ versus $g(\Delta f) - r(\Delta f)$ colour-magnitude
diagram for us to contemplate using any such cut. The detailed form of
the cut in equation~(\ref{eq:cmdcuts}) is chosen on the basis of
numerical simulations of the source population, which suggest that the
sources lie overwhelmingly in the band $0< R(\Delta f)-I(\Delta
f)<1.5$ (see for example Figure~\ref{fig:eamonn} discussed later).

The {\bf seventh cut} is used to distinguish convincing, or
first-level, candidates from more tentative, or second-level,
candidates. First, an estimate of the signal-to-noise ratio in
microlensing $(S/N)_{\rm micro}$ is computed using
\begin{equation}
\left( S/N \right) = \sqrt{ {1\over N} \sum_i \left[ {f_i - f_{\rm bl}
\over \sigma_i} \right ]^2},
\label{eq:sn}
\end{equation}
where the sum is over all $N$ datapoints within \tfwhm, while $f_i$ is
the flux measurement, $\sigma_i$ is its error and $f_{\rm bl}$ is the
baseline of the global microlensing fit.  Then, an estimate of the
signal-to-noise ratio in the residuals $(S/N)_{\rm res}$ is calculated
as follows. The median of $|f_i - f_{\rm bl}|$ is calculated for the
set of flux measurements outside 3 \tfwhm\ of either side of the
peak. Then, equation~(\ref{eq:sn}) is applied to compute $(S/N)_{\rm
res}$, with the sum running over all datapoints above the median (and
outside the event).  These signal-to-noise ratios are computed using
all the data in each of the three passbands ($g,r$ and $i$). As
Figure~\ref{fig:ston} shows, there is a cloud of datapoints in the
($(S/N)_{\rm micro}$,$(S/N)_{\rm res}$) plane, together with a clear
tail of events with high $(S/N)_{\rm micro}$. The cut
\begin{equation}
\left ( S/N \right)_{\rm micro} > 15 + \left( S/N \right)_{\rm res}
\label{eq:stona}
\end{equation}
is applied to leave \final first-level microlensing candidates. The
underlying meaning of this cut is that the signal-to-noise ratio in
microlensing must exceed the signal-to-noise ratio in the
residuals. Although the precise form of equation~(\ref{eq:stona}) is
somewhat arbitrary, there is nonetheless a very clear separation
between the \final high signal-to-noise events and the rest of the
datapoints on Figure~\ref{fig:ston}.  It is also useful to extract a
looser set of candidates, whose microlensing nature is possible but
less certain.  The cuts
\begin{equation}
\left ( S/N \right)_{\rm micro} > 4 + \left( S/N \right)_{\rm res},
\qquad\qquad  \left( S/N \right)_{\rm res} < 2,
\label{eq:stonb}
\end{equation}
are used to isolate a further \finals microlensing candidates, which we
refer to as second-level.

Finally, it also interesting to examine the lightcurves just below our
threshold for designation as microlensing candidates. Accordingly, we
extract a set of \finalt third-level lightcurves with the cuts:
\begin{equation}
\left ( S/N \right)_{\rm micro} > 4 + \left( S/N \right)_{\rm res},
\qquad\qquad  \left( S/N \right)_{\rm res} > 2.
\label{eq:stonc}
\end{equation}
As we will see, these are almost certainly mostly variable stars.

\begin{figure*}
\includegraphics[width=\hsize,height=21cm]{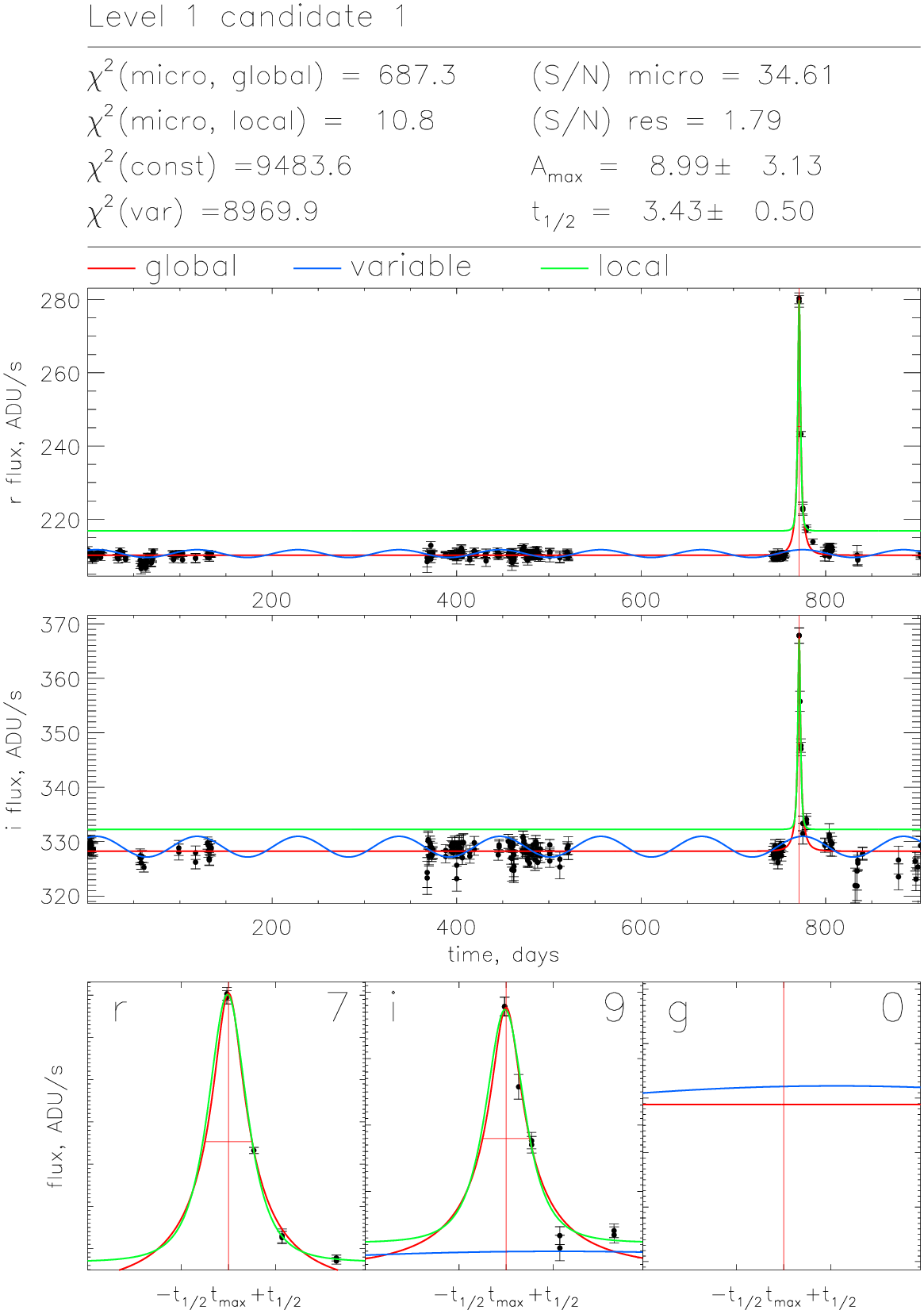}
\caption{Lightcurves in the $r$ and $i$ bands of 
first-level candidate 1 in Table~\ref{tab:params} from the southern
field, CCD 2.  Overplotted in blue is the best bumpy variable fit, in
red is the best global microlensing fit and in green is the best local
microlensing fit. The location of the peak of the event is marked by a
red vertical line. The lower sub-panels show the peak in $g$, $r$ and
$i$, and the number of datapoints is recorded in the top corner of
each sub-panel.}  
\label{fig:good1} 
\end{figure*} 
\begin{figure*}
\includegraphics[width=\hsize,height=21cm]{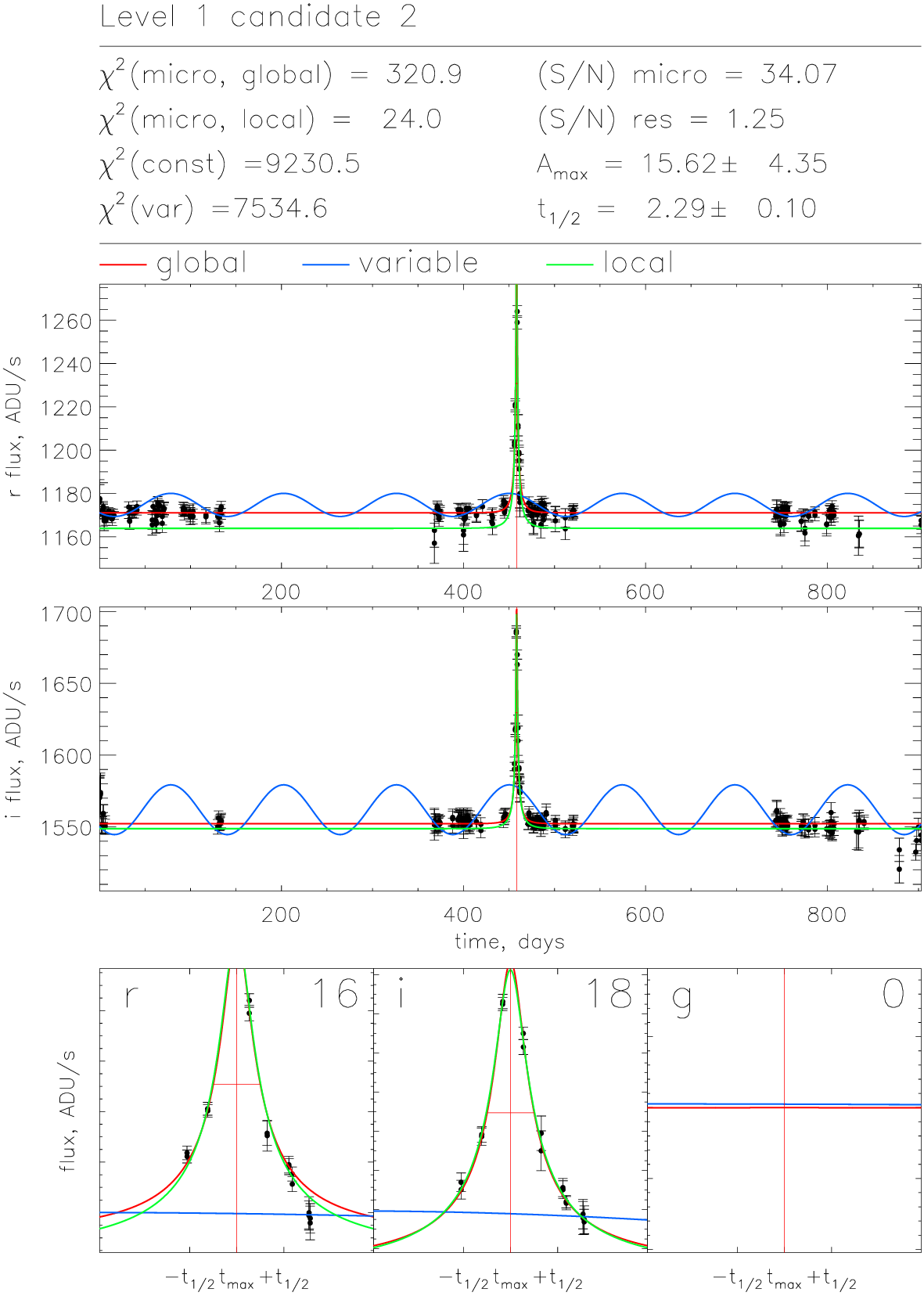}
\caption{Lightcurve of first-level candidate 2 in
Table~\ref{tab:params} from the southern field, CCD 3.
This is PA 00-S3 of Paulin-Henriksson et al. (2003) and
WeCAPP-GL-1 of Riffeser et al. (2003).}
\label{fig:good2}
\end{figure*}
\begin{figure*}
\includegraphics[width=\hsize,height=21cm]{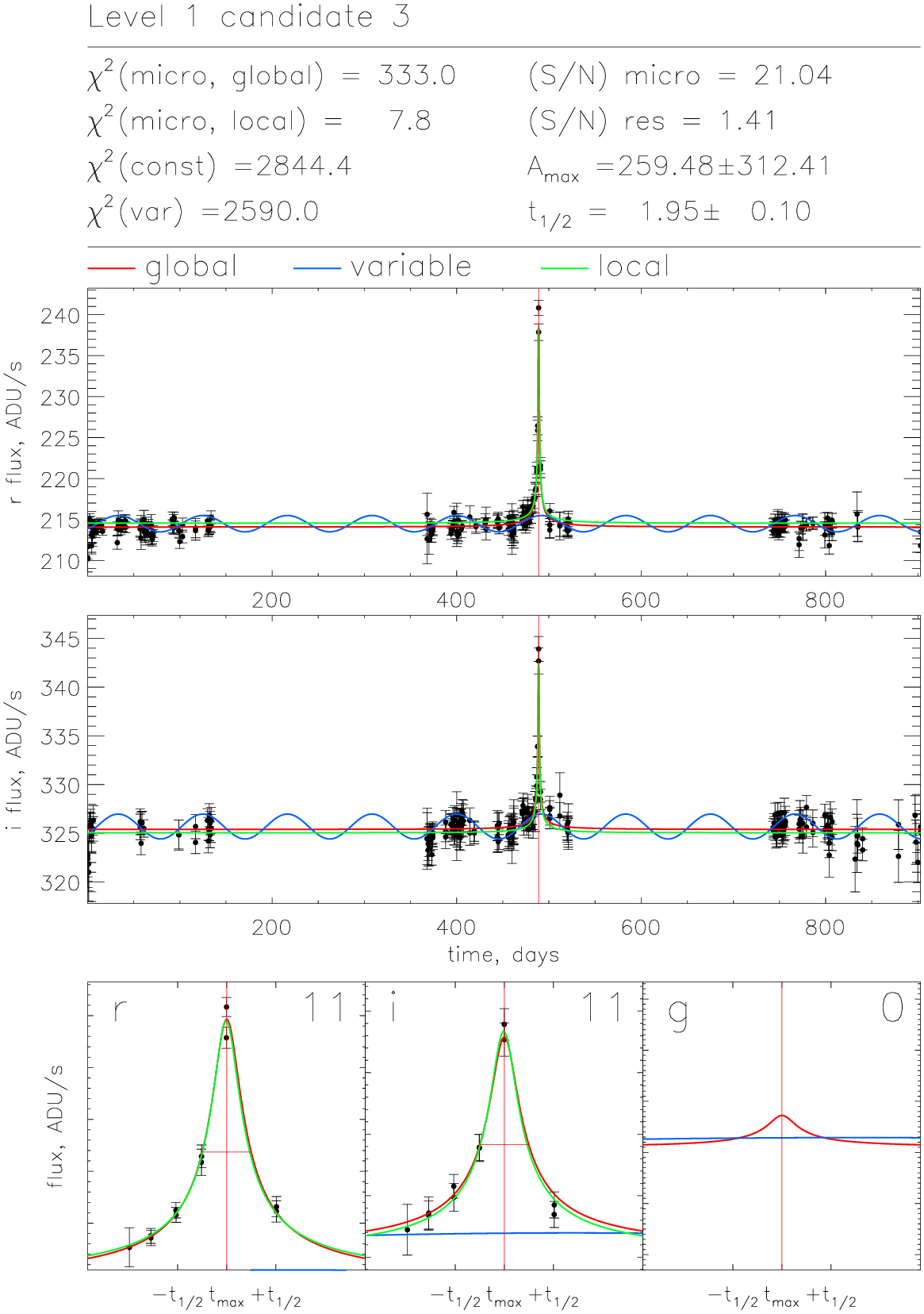}
\caption{Lightcurve of first-level candidate 3 in
Table~\ref{tab:params} from the southern field, CCD 4.
This is PA 00-S4 of Paulin-Henriksson et al. (2003).}
\label{fig:good3}
\end{figure*}

\begin{table} 
\begin{center}
\begin{tabular}{ccccccc}
\hline
Candidate & Level & Field/CCD & R & RA & Dec \\
\hline
1 & 1& F2,CCD2& 22\farcm94& 00\h42\m02\fs3& 40\degr54\arcmin35\arcsec\\
2 & 1& F2,CCD3&  4\farcm08& 00\h42\m30\fs3& 41\degr13\arcmin01\arcsec\\
3 & 1& F2,CCD4& 22\farcm54& 00\h42\m30\fs0& 40\degr53\arcmin46\arcsec\\
\hline
1 & 2& F2,CCD1& 30\farcm64& 00\h42\m57\fs7& 40\degr45\arcmin37\arcsec \\
2 & 2& F2,CCD3& 3\farcm40 & 00\h42\m59\fs5& 41\degr14\arcmin17\arcsec \\
3 & 2& F2,CCD3& 5\farcm57 & 00\h42\m23\fs9& 41\degr12\arcmin06\arcsec \\
\hline
\end{tabular}
\end{center}
\caption{The locations of the \final first-level and the \finals second-level
candidates in right ascension and declination (J2000.0), together with
the projected distance $R$ from the centre of
M31.}
\label{tab:locs} 
\end{table}

\begin{table*} 
\begin{center}
\begin{tabular}{cccccccccc}
\hline
Candidates & Level & $t_{\rm max}$ & \tfwhm & $\tE$ & $A_{\rm max}$ & $f_{s,r}$
& $f_{b,r}$ & $f_{s,i}$ & $f_{b,i}$\\
\hline
 1&  1 & $771.3 \pm  0.2$&  $  3.4 \pm  0.5$&     $10.1 \pm    1.8$&  $  8.99 \pm   3.128$&  $  
8.8 \pm    3.1$&  $ 201.4 \pm    3.1$&  $   4.9 \pm    1.8$&  $ 323.3 \pm    1.8$ \\
 2&  1& $458.4 \pm  0.0$&  $  2.3 \pm  0.1$&     $11.2\pm    2.6$&  $ 15.62 \pm   4.350$&  $  
8.2 \pm    2.3$&  $1162.9 \pm    2.2$&  $  10.6 \pm    3.0$&  $1541.6 \pm    2.9$ \\
 3&  1 & $488.9 \pm  0.1$&  $  2.0 \pm  0.1$&    $146.6 \pm  174.2$&  $259.48 \pm 312.411$&  $  
0.1 \pm    0.1$&  $ 214.0 \pm    0.1$&  $   0.1 \pm    0.1$&  $ 325.3
\pm    0.1$ \\ \hline
1& 2&  $753.6 \pm  1.2$&  $ 35.8 \pm  3.8$&  $   36.3 \pm   28.0$&  $  2.57 \pm   2.407$&  $   3.2
 \pm    4.9$&  $ 132.4 \pm    4.9$&  $   2.2 \pm    3.4$&  $ 230.4 \pm
 3.4$ \\
 2& 2&  $ 46.6 \pm  0.5$&  $ 33.1 \pm  2.1$&  $   11.1 \pm    1.4$&  $  1.06 \pm   0.032$&  $ 257.9
 \pm  127.4$&  $ 678.6 \pm  127.4$&  $-$&  $-$ \\
 3&  2& $130.7 \pm  2.9$&  $ 50.7 \pm  5.4$&  $   21.4 \pm    5.2$&  $  1.15 \pm   0.122$&  $ 102.3
 \pm   84.5$&  $ 711.5 \pm   84.5$&  $ 295.8 \pm  243.2$&  $ 797.6 \pm
 243.2$  \\ \hline
\end{tabular}
\end{center}
\caption{The parameters of the microlensing fit for the \final
first-level and the \finals second-level candidates: $t_{\rm max}$ is
the time of the peak measured in days from JD 2451392.5, \tfwhm\ is
the full-width half-maximum timescale in days, $\tE$ is the Einstein
crossing time, $A_{\rm max}$ is the maximum amplification and $f_s,
f_b$ are the source and blend fluxes in ADU s$^{-1}$ in the $r$ and
$i$ bands. Notes: [1] the error bars assume symmetric Gaussian
distributions; [2] Level 2, candidate 2 has no signal in the $i$
band.}  
\label{tab:params} 
\end{table*}

\subsection{Properties of the Candidate Selection Algorithm}

Table~\ref{tab:events} records the total number of lightcurves
surviving each successive cut in the candidate selection algorithm,
broken down according to the field and CCD combinations.  Note that
genuine microlensing events with very short $\tfwhm$ may have failed
to pass the first cut.  For example, if a short-timescale event is
superposed on a bumpy baseline (caused by variable stars contributions
to the same superpixel), then the comparatively small number of
datapoints in the microlensing event itself are outnumbered by the
larger number of datapoints participating in the overall periodicity
of the stellar variable. Such a lightcurve is better fitted by a
variable star template rather than a microlensing one and so such
short-timescale events fail the first cut. Similarly, the fifth cut
(achromaticity) and the sixth cut (colour-magnitude) are especially
needed for long-timescale events for which we cannot rely on
periodicity to eliminate all repeating variables.

Figure~\ref{fig:cuts} shows the effect of each of the cuts applied
singly to the starting catalogue of \total lightcurves on the
distributions of colour $R(\Delta f)-I(\Delta f)$ and magnitude
$I(\Delta f)$. With the exception of the initial $\Delta \chi ^2$ cut,
which tends to promote brighter events, there is no obvious bias in
magnitude introduced by the first six cuts. The same holds true for
the colour distributions, with the exception of the explicit
truncation introduced by the requirement $R(\Delta f) - I(\Delta f) <
1.5$ in the sixth cut.

This cut in the colour-magnitude diagram merits further discussion.
By examination of Figure~\ref{fig:cmds}, it is apparent that the cut
is very loose. For example, another possibility is to excise the
entire area in the colour-magnitude diagram that is occupied by Miras
and long-period variables, by choosing cuts that tightly wrap around
the densest regions.  We elected not to do this, as this region of the
colour-magnitude diagram is also where many of the potential
microlensing sources are located.

A simple test of the plausibility of the magnitude and colour of our
\final first-level candidates is to see how representative they are of
their local potential microlensing source population. By testing for
local rather than global consistency, we avoid having to assume some
underlying microlensing model. We define the potential source
population by a simple $S/N$ criterion that reflects the quality of
our candidate events without requiring a detailed simulation of the
effect of each cut in our pipeline. If the cuts in our pipeline impose
a strong bias upon the colour or magnitude of selected events, then we
can not expect a good agreement with a local source population
selected by a simple S/N cut. Similarly, there is no reason to expect
agreement if the candidates are variable stars rather than
microlensing events.

The sources are drawn from synthetic stellar population models
(c.f. Girardi \& Salaris 2001), assuming similar mass, age and
metallicity distributions to the Milky Way bulge and disk.  For each
synthetic star we compute the maximum impact parameter $u_{\rm max}$
that permits a detection of the magnification peak with $S/N =
20$. Since $u_{\rm max}$ depends upon the local M31 surface brightness,
it is computed for the location of each of the three candidates using
the surface photometry of Walterbos \& Kennicutt (1987). Since the
microlensing rate scales with $u_{\rm max}$, this parameter is
essentially the statistical weight that the star may act as a
potential microlensing source.

Figure~\ref{fig:eamonn} shows the transformed Cousins
$I(\Delta f)$ versus $R(\Delta f)-I(\Delta f)$ pseudo-magnitudes for
each of the \final first-level candidates, plotted together with their
respective local source populations, as selected by their $u_{\rm
max}$ weighting. The pseudo-magnitude for each synthetic star is
computed from a randomly-realized impact parameter $u$ between 0 and
$u_{\rm max}$. For candidates 1 and 3 we employ a disk synthetic
stellar population, whilst for candidate 2 we use a bulge stellar
population model. The positions of candidates 1 and 3 appear
consistent with main-sequence disk source stars, whilst candidate 2 is
consistent with a lensed bulge giant source star. The consistency
between the magnitudes and colours of the three events and their local
source populations bolsters their interpretation as microlensing
events.

\begin{figure}
\includegraphics[height=13cm]{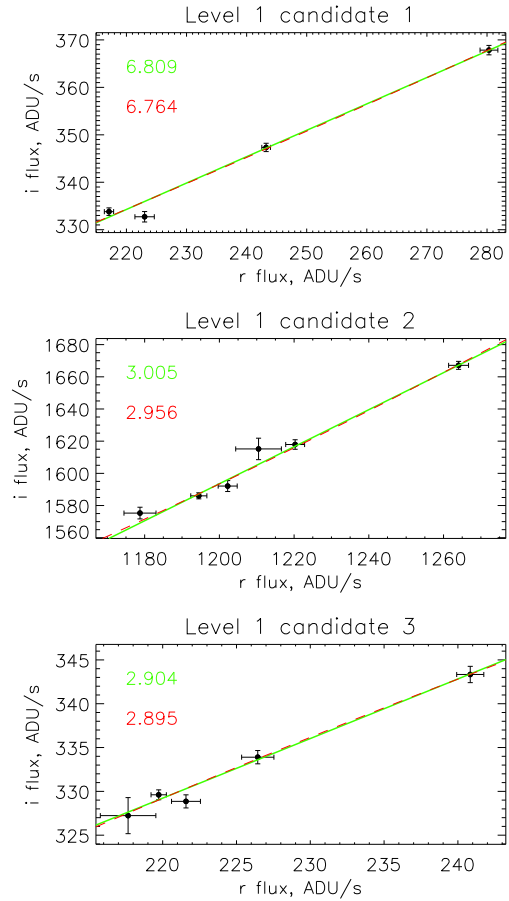}
\caption{Achromaticity curves for the \final first-level
candidates. The flux in $r$ is plotted against the flux in $i$,
together with the best-fitting straight line (solid green) and
quadratic curve (dashed red). The numbers in green (red) in the top
corners of each panel are the $\chi^2$ values for the linear
(quadratic) fits.}
\label{fig:chromo}
\end{figure}

\section{The \final First-Level Candidates} 

\subsection{The Lightcurves}

Figures~\ref{fig:good1}-\ref{fig:good3} show three-year lightcurves of
the \final first-level candidates in the $r$ and $i$ bands, together
with close-ups of the peaks in $g, r$ and $i$, if data are available.
Overplotted on all the lightcurves are the best variable star fit (in
blue), the best global microlensing fit (in red) and the best local
microlensing fit (in green). The time of maximum magnification $t_{\rm
max}$ is indicated by a red vertical line and the full-width
half-maximum timescale $\tfwhm$ by a red horizontal line. The lower
three sub-panels show blow-ups of the datapoints within 1.5 \tfwhm\ on
either side of the peak. These are the datapoints used in the local
microlensing fit. The number of datapoints in each passband is
recorded in the top right-hand corner.  The $\chi^2$ values calculated
in the selection algorithm, together with the signal-to-noise ratios,
are recorded in the top panel of each lightcurve, enabling the reader
to locate the positions of the individual events on
Figures~\ref{fig:chisq} and \ref{fig:ston}.  Figure~\ref{fig:chromo}
shows the achromaticity test for the \final first-level candidates.
The red (green) curves show the straight line (quadratic) fit to the
flux measurements in two passbands. In all cases, the quadratic
coefficients are redundant, as judged by the $\chi^2$ given in the top
corners of the panels of Figure~\ref{fig:chromo}.

\begin{figure*}
\vspace{6cm}\centering{\tt 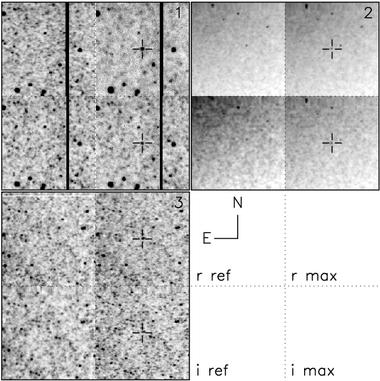}\vspace{6cm}
\caption{Thumbnail images ($1' \times 1'$) of candidates 1 (top left),
2 (PA 00-S3, top right) and 3 (PA 00-S4, bottom). Each panel has 4
sub-panels. In each panel, the $r$ band is at the top, the $i$ band at
the bottom, while left corresponds to the reference frame and right to
the peak epoch. The location of the event is always at the centre of
the frame and marked with a cross in the right panels.}
\label{fig:thumbs} 
\end{figure*}

Table~\ref{tab:locs} gives the locations of the \final first-level
candidates in right ascension and declination, together with the
projected distance from the centre of M31.  All lie in the southern
field, one each on CCDs 2, 3 and 4.  Table~\ref{tab:params} shows the
parameters of the microlensing fit for the \final candidates. Nine
microlensing parameters were fitted, namely the source flux in $g, r$
and $i$, the blending flux in $g, r$ and $i$, the impact parameter,
the time of maximum and the Einstein radius crossing time) using all
datapoints in all bands. However, the $g$ band data are patchy, so the
source and blend fluxes in $g$ are poorly constrained and not reported
in Table~\ref{tab:params}. All 3 events have magnification $A_{\rm
max}$ above the threshold value of $3/ \sqrt{5}$ = 1.34 and so have
impact parameter smaller than unity. All 3 have very short timescales
($\tfwhm = 2.0, 2.3$ and 3.4 days respectively). For each of the
\final candidates, thumbnail $r$ and $i$ band images of the reference
frame (left) and the frame of the peak of the event (right) are shown
in Figure~\ref{fig:thumbs}.

\subsection{Candidate 1}

Candidates 2 and 3 have been discussed previously by Paulin-Henriksson
et al. (2002, 2003). However, candidate 1 is new and worthy of some
detailed discussion here.

While candidate 1 has formally passed the second cut, we find that,
from Figure~\ref{fig:good1}, the rising part of the lightcurve is not
well constrained from the available data. This raises the question as
to whether this lightcurve can be modelled by certain types of
cataclysmic variable, notably a nova. This is particularly worrisome
as the calibrated colour of the candidate ($R-I=-0.1$) is close to
that of the classical novae found in the POINT-AGAPE dataset by
Darnley et al. (2004). Furthermore, a fair fraction of the novae of
Darnley et al. (2004) do not show very significant chromatic behaviour
over a short timescale. However, we find that the observed magnitude
at the maximum of the light curve ($R_{\max}\la 19.5$) is rather faint
compared to the other novae of Darnley et al. (2004) and that the
decline rate is rather too fast ($dR/dt\ga 0.5~\mbox{day}^{-1}$),
although the sparse sampling may mean that we missed the actual
maximum, which may have occurred earlier and been brighter. The
possibility of this being a nearby dwarf nova is much less
significant, as the lightcurve shows no sign of repeating bursts. It
is also very unlikely to be a background (super-)nova considering its
fairly blue colour.  Its presence in the disk strongly suggests that,
if it were a background event, it would be heavily reddened.

In fact, the microlensing fit result for candidate 1 indicates that
the source should be relatively bright if this is indeed a
microlensing event. Based on equation~(\ref{eq:bloodylimit}), we find
the nominal detection limit of the source in the $r$ band at the
location of candidate 1 to be $5.8\ \mbox{ADU s$^{-1}$}$, while the
source flux from the fit is $8.8\pm3.1\ \mbox{ADU s$^{-1}$}$. In other
words, although this candidate has passed the fourth cut, it is still
just possible that we may find a resolved source at the reference
frame, if the microlensing fit is correct\footnote{This is not in
contradiction with the fact that we masked the region around the
`resolved' stars as the mask was around a \emph{bright} resolved
star. The typical magnitude of the resolved stars used in building the
mask is $R\sim20.5$. The source magnitude corresponding to the fit
result of candidate 1 is much fainter at $R=21.9\pm0.4$.}.

After examining the frames when the event is at baseline, including
the reference frame, we find `a resolved object' near the event ($\la
1$ pixel).  The fixed aperture (10 pixel aperture) photometry on this
object indicates that, although the $r$ band flux ($9.5\pm1.7$ ADU
$s^{-1}$) is consistent with this being the microlensed source, the
$i$ band flux ($12.1\pm2.0$ ADU $s^{-1}$) is rather too bright --
i.e., `the colour' of this object is too red. However, we note that it
is very likely that, for given seeing, most `resolved objects' in our
images are blended mixtures of several stars and thus it is still
possible that the object actually contains the microlensed
source. This idea could be tested by examining an image with better
resolution, such as a {\it Hubble Space Telescope} (HST)
image. Unfortunately, the nearest HST archival pointing is 2.6 arcmin
away. Another possibility is to examine the precise position of the
centroid of this object and show that, while the $r$ and $i$ band
centroid are shifted, the $r$ band position is closer to the centroid
position of the event at maximum. This, though, is a difficult
calculation to perform convincingly, as it requires precise modelling
of the point spread function.

If we accept the fit, the magnitude and the colour of the source
($R=21.9\pm0.4$, $R-I=-0.1$) are consistent with that of an M31 main
sequence B star under moderate extinction. This is quite reasonable as
the event occurs on the disk near the dust lane, possibly along the
star formation region or close to an OB association.

\begin{figure}
\vspace{6cm}\centering{\tt 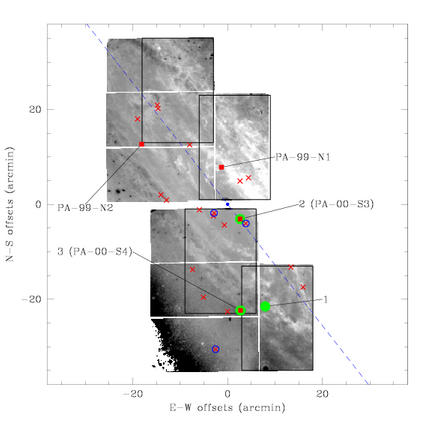}\vspace{6cm}
\caption{The locations of our \final first-level candidates (green
circles) are superposed on a $g-r$ greyscale map of M31.  Red squares
are the locations of the 4 candidates given in Paulin-Henriksson et
al.  (2003). Also shown are the \finals second-level candidates (red
crosses within blue circles) and the \finalt third-level lightcurves
(red crosses).  The blue line marks the projected major axis and so
separates the near and far disk. The optical centre of M31 is marked
with a blue dot.  The black rectangles indicate symmetric portions of
M31 defined by An et al. (2004b). Combining all the first-level,
second-level and third-level candidates, there are 15 events in the
symmetric regions of the INT fields.}
\label{fig:m31map}
\end{figure}

\subsection{Comparison with Other High Signal-to-Noise Surveys}

Paulin-Henriksson et al. (2003) carried out a survey fine-tuned for
short-timescale ($\tfwhm < 25$ days), high signal-to-noise ratio
($R(\Delta f) < 21$) microlensing events using the data taken in the
1999 and 2000 seasons. They found 4 events. One of these (PA 99-N2) is
missing from our starting catalogue of \total variable lightcurves as
it has been masked out. This very bright event by chance occurs on the
reference frame used to construct the resolved star catalogue (see An
et al. 2004b). It is so bright that it is identified by our software
as a resolved star!  Hence, the area around PA 99-N2 is removed in the
masking. An elaboration of our procedure -- in which two resolved star
catalogues are constructed at different epochs, and objects present in
only one are identified as variables and so not deleted -- would
presumably allow us to detect PA 99-N2. A further one of the
Paulin-Henriksson et al. (2003) candidates (PA 99-N1) is missing
because it fails the first cut of our selection algorithm. It is a
short-timescale candidate ($\tfwhm = 1.8$) with a noisy baseline. It
was originally selected based on a single significant peak in the 1999
data. Closer analysis over a longer timeframe, however, revealed that
the baseline of the pixel lightcurve was bumpy, and that this was
caused by contamination from a bright variable star located $1\farcs1$
south of the microlensed source (Auri\`ere et al. 2001). The remaining
two of the Paulin-Henriksson et al. (2003) candidates (PA 00-S3 and PA
00-S4) are present in our sample of first-level candidates. 

Very recently, a further 3 short-timescale events have been detected
by Paulin-Henriksson et al. (2004) using a different and independent
candidate selection algorithm to this paper. A detailed analysis of
these events is in preparation (Calchi Novati et al. 2004).

\begin{table}
\begin{center}
\begin{tabular}{cc}
\hline
MEGA candidate & Cuts Failed \\
\hline
3 & 2, 7 (level 1) \\ 
5 & 1, 2, 3, 7 (all levels)\\
8 & 6, 7 (all levels) \\
9 & 2, 7 (all levels) \\
10 & 7 (all levels) \\
11 & - \\
\hline
\end{tabular}
\end{center}
\caption{The MEGA candidates present in our starting catalogue and the
cuts in our pipeline that they fail. Note that MEGA 11 is PA 00-S4 of
Paulin-Henriksson et al. (2003) and is one of the first-level
candidates in this paper.}
\label{tab:MEGA}
\end{table}

The MEGA collaboration (de Jong et al. 2004) presented 14 candidate
events based on the same INT imaging data as this paper, but using a
different data analysis pipeline and candidate selection algorithm.
Of these 14 events, 12 were new and 2 (PA 99-N2 and PA 00-S4) had
already been discovered by Paulin-Henriksson et al. (2003). By
matching the pixel positions, we find that (at most) 6 of the MEGA
candidates are present in our catalogue of \total lightcurves, namely
MEGA 3, 5, 8, 9, 10 and 11. MEGA 1, 2, 4, 6, 12 and 13 are not present
even in the catalogue before masking, as they are not judged to have 3
consecutive points above 3 $\sigma$.  MEGA 7 (PA 99-N2) and 14 are
removed by our mask.  For the remaining six candidates,
Table~\ref{tab:MEGA} gives the cuts in our pipeline that are
failed. Note that MEGA 11 is the same as PA 00-S4, and so lies in our
set of first level candidates. Apart from MEGA 11, the remaining
candidates all fail at least one of our cuts.  MEGA 10 only fails the
final signal-to-noise ratio cut, and in fact lies very close to the
second-level boundary in our Figure~\ref{fig:ston}.  MEGA 3 would be
adjudged a second-level candidate were it not for the fact that it
failed our second cut on sampling.

\subsection{The Asymmetry Signal}

The spatial locations of our \final first-level candidates, as well as
the 4 candidates of Paulin-Henriksson et al. (2003), are plotted on a
$g-r$ greyscale map of M31 in Figure~\ref{fig:m31map}. There are 2
events in common, so in total this gives 5 distinct candidates.  Also
shown is the projected major axis in blue. One of the
Paulin-Henriksson et al.  candidates (PA 99-N1) lies in the near-side
of the disk; all of the remaining events occur in the
far-side. Notwithstanding the claim of de Jong et al. (2003), it is
hard to identify spatial gradients in the microlensing rate by ocular
examination of such maps. As we have emphasised elsewhere (An et
al. 2004b), the near-far asymmetry signal is confused by the effects
of variable extinction and dust associated with the M31 spiral
arms. This is a serious problem because the dust distribution is
asymmetric over a large scale and, even worse, the dust asymmetry is
maximised roughly along the major axis and so mimics the expected
signal from microlensing by halo dark matter.  This led An et
al. (2004b) to propose searching for an east-west asymmetry, as the
variable star distributions are roughly symmetric about the
north-south line through the M31 centre. Considering the entire
sample, the division is 2 events east and 3 events west of the
north-south line.  The very small number of convincing microlensing
candidates that have unambiguously been determined from the three
years of M31 imaging implies that such ratios are dominated by Poisson
noise.

Note that absence of an evident asymmetry may still yield a strong
result on the baryon fraction of the M31 halo, as many models predict
asymmetries that should be easily detectable after three-years data
collection (see Kerins et al. 2001). We will return to this matter in
a later paper, in which the efficiency of our survey will be
calculated.

\begin{figure*}
\includegraphics[width=\hsize,height=21cm]{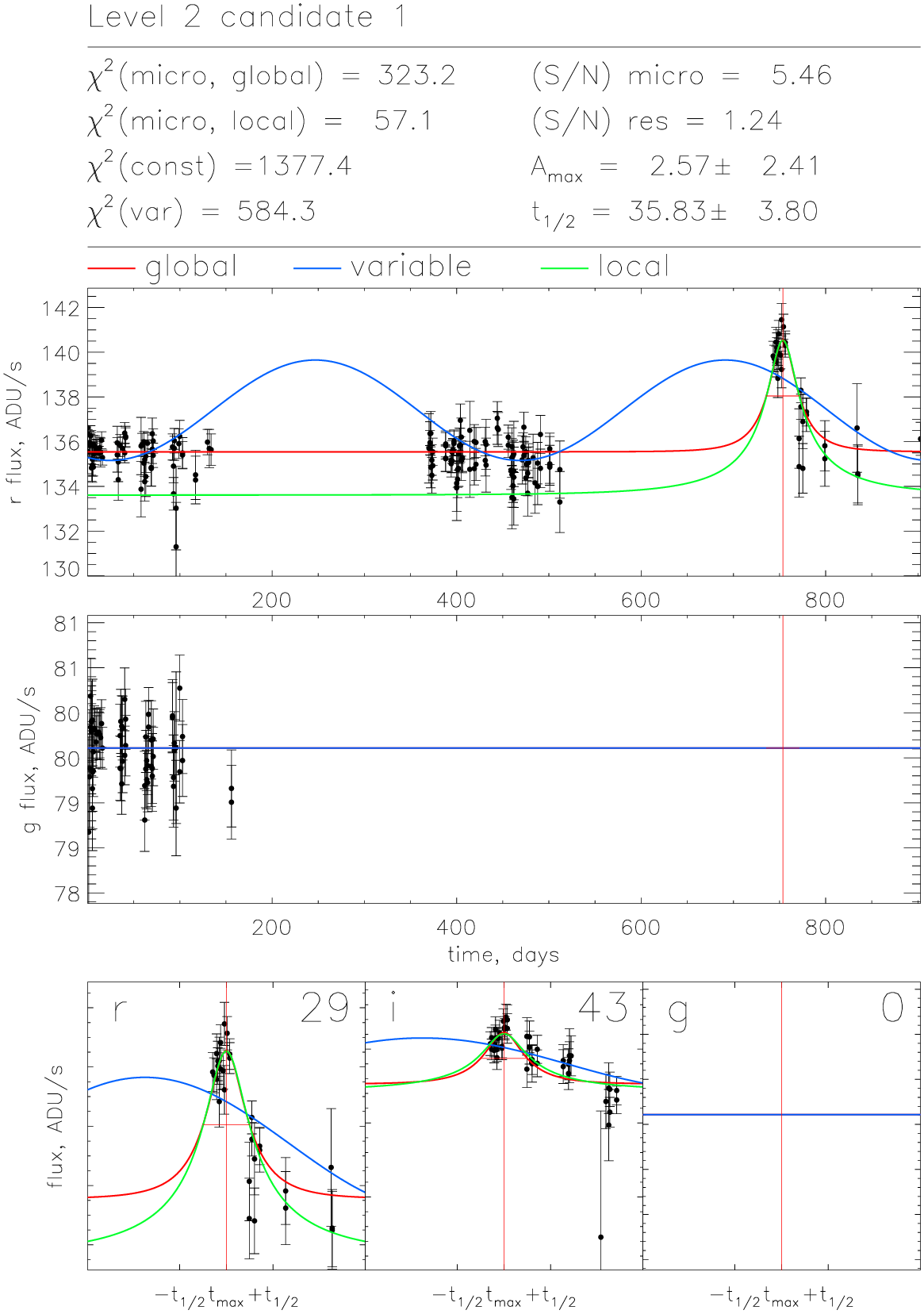}
\caption{Lightcurves in the $r$ and $i$ bands of second-level
candidate 1 in Table~\ref{tab:params} from the southern field, CCD 1
Overplotted in blue is the best bumpy variable fit, in red is the best
global microlensing fit and in green is the best local microlensing
fit. The location of the peak of the event is marked by a red vertical 
line. The lower sub-panels show the peak in $g$, $r$ and
$i$, and the number of datapoints is recorded in the top corner of each
sub-panel.}
\label{fig:bad1}
\end{figure*}
\begin{figure*}
\includegraphics[width=\hsize,height=21cm]{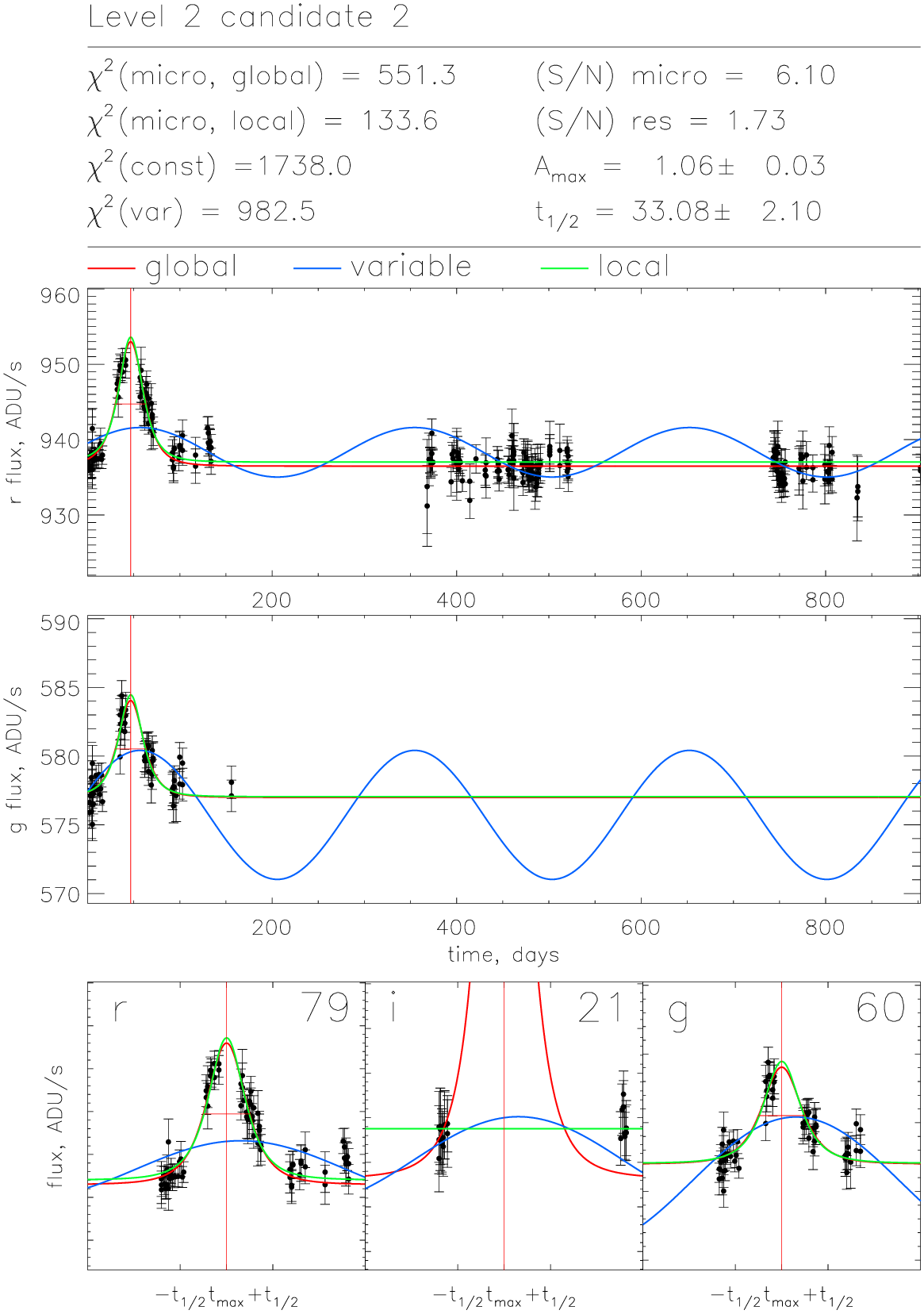}
\caption{Lightcurves in the $r$ and $g$ band of second-level 
candidate 2 in Table~\ref{tab:params} from the southern field, CCD 3.}
\label{fig:bad2}
\end{figure*}
\begin{figure*}
\includegraphics[width=\hsize,height=21cm]{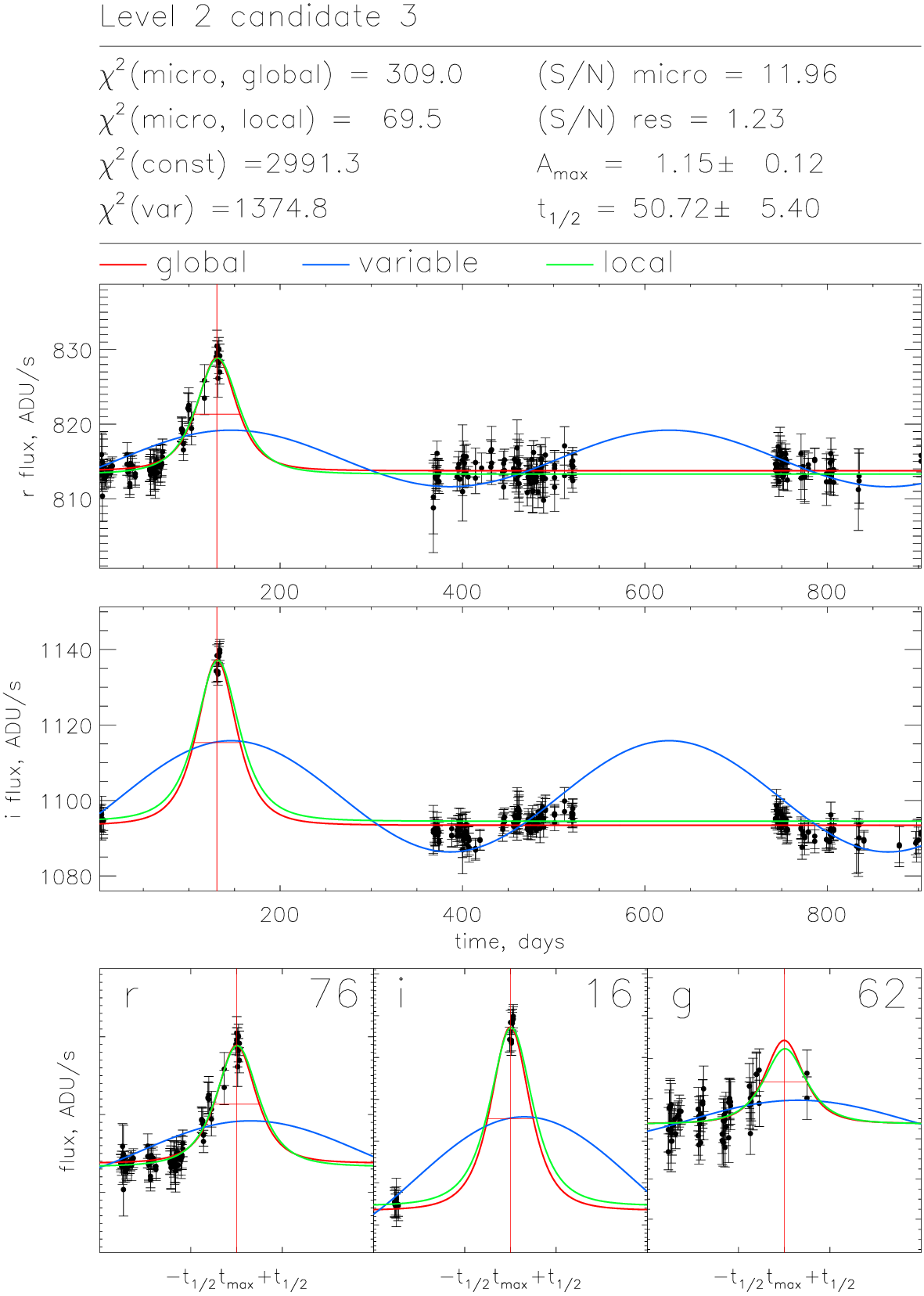}
\caption{Lightcurve of second-level candidate 3 in
Table~\ref{tab:params} from the southern field, CCD 3.}
\label{fig:bad3}
\end{figure*}

\section{The \finals Second-Level Candidates and the 
\finalt Third-Level Lightcurves}

Bearing in mind that the second-level candidates are less securely
established as microlensing, we briefly discuss some of their
properties.  The lightcurves of the \finals second-level candidates
are shown in Figures~\ref{fig:bad1}-\ref{fig:bad3}.  The parameters of
the microlensing fits and the event locations are given in
Tables~\ref{tab:locs} and \ref{tab:params}.  On scrutinising the
lightcurves, it is immediately apparent that -- although the
microlensing interpretation is possible -- the lightcurves are in
general noisier and sometimes show clear evidence of additional
structure.  The additional structure may be evidence of contamination
from other nearby variable sources, or it may be evidence that
some of the lightcurves do not correspond to genuine microlensing
events at all. All \finals second-level candidates need to be examined
over a longer baseline before they can be accepted or rejected as
genuine microlensing.

It is also interesting to see the lightcurves that lie just below our
threshold for designation as microlensing candidates. These are the
\finalt third-level lightcurves, which are displayed compactly in
Figure~\ref{fig:vas1} of the Appendix B. The lightcurves are either
those of variable stars or are highly contaminated by nearby variable
stars. There are also indications that -- here and there -- our
fitting software has not always found the best variable star fit to
the datapoints. So, the lightcurves have survived to the final or
seventh cut in our selection algorithm because of deficiencies in our
comparison between the variable star and global microlensing
templates.  The locations of the events are listed in
Table~\ref{tab:locsthree}. Eight of the 16 third-level events occur in
the first season, 2 in the second and 6 in the third. The second
season, which has the largest number of epochs, contains the fewest
events. This suggests very strongly that some (probably most) of the
candidates are variables and that poor sampling is playing a role in
allowing rogue candidates to survive the cuts.  If the bump of a
long-period variable occurs in the second season, then the end of the
fall in the first season, and the beginning of the rise in the third
season can be sampled, and so the event is discarded as possible
microlensing.  If the bump occurs in the first season, then the second
bump can lie in the $\sim 6$ month gap in sampling after the second
season, when M31 is not visible from La Palma, and so the lightcurve
can survive. In the first season, data were taken predominantly in the
$g$ and $r$ bands, whilst in the second and third season predominantly
$r$ and $i$ band data were taken. For these red variables, a
well-sampled lightcurve in the $i$ band is best for uncovering
discrepancies from the blended Paczy\'nski fits. This pattern of
sampling in different bands is almost certainly the explanation why
half of all the third-level lightcurves peak in the first season.
Although there may be one or two genuine microlensing candidates in
the third-level sample, they are almost certainly mostly variable
stars. We therefore do not dignify them as microlensing candidates.

There are a total of 22 first-level, second-level and third-level
lightcurves in all.  Figure~\ref{fig:cmd20} shows the entire sample on
a colour-magnitude diagram (except for two events for which there is
no $i$ band signal), while Figure~\ref{fig:m31map} shows their
locations in M31.
\begin{figure}
\vspace{0.4\hsize}\centering{\tt fig16.gif}\vspace{0.4\hsize}
\caption{The locations of 20 events in the colour-magnitude
diagram. The first-level candidates are denoted by green dots, the
second-level candidates by red crosses within blue circles and
third-level lightcurves by red crosses. (There are 2 lightcurves for
which there is no $i$ band signal. Note too that the locations of the
candidates are deduced using the microlensing fit, and so differ
slightly from the coresponding locations in Fig.~3.)}
\label{fig:cmd20}
\end{figure}
Seventeen of the 22 events lie within the symmetric regions of the INT
fields. We can use the asymmetry signal-to-noise ratio definition of
Kerins et al. (2003) on these 17 events to check the significance of
their distribution.  We find that the near-far and east-west signals
are well within Poisson error, whereas the north-south signal appears
to be marginally significant. From An et al. (2004b), we recollect
that a north-south signal is consistent with a population of objects
affected by dust since the $g-r$ image of M31 indicates more dust in
the north symmetric region.  So, regardless of whether our
second-level and third-level events are predominantly microlenses or
variable stars, the significant north-south asymmetry indicates that
their distribution appears to be affected by the dust distribution
across the M31 disk. The lack of far-near or east-west asymmetry again
argues against a microlensing hypothesis for most of these
events. 

Alcock et al. (2000) have already established a precedent in
microlensing studies by publishing a set of strong candidates (their
``set A'') and a set of weaker candidates (their ``set B'').  The
advantage of this is that the optical depth can then be studied as a
function of the threshold of detection.  We have followed their
example by defining a set of first-level (or strong) candidates and
second-level (or weaker) candidates.  A detailed comparison of the
numbers of first-level and second-level microlensing candidates with
theoretical predictions requires a careful calibration of our
detection efficiency as a function of threshold and will be presented
in a follow-up paper.

\section{Conclusions} 

The POINT-AGAPE collaboration has presented the results of a fully
automated search through a catalogue of \total variable lightcurves
built using the superpixel method from three years of imaging data
towards M31. In order to establish that the results obtained from
microlensing experiments are robust, one should compare candidate
samples derived from different sets of selection criteria.  Our
selection algorithm has no explicit restriction on timescale.  It is
thus complementary to the work of Paulin-Henriksson et al. (2003,
2004) and Calchi Novati et al. (2004), who use a different algorithm
that is fine-tuned for short, high-amplitude events, and to Tsapras et
al. (2004) who apply another, independent set of selection criteria.
At the end of our automated search, we obtained a set of \final
first-level candidates, which are the most convincing. All these
events have short full-width half-maximum timescales. We also obtained
a set of \finals second-level candidates, for which microlensing is a
possible hypothesis.

We regard the \final first-level lightcurves that pass all stages of
the pipeline as excellent microlensing candidates. This includes the
events PA 00-S3 and PA 00-S4 identified by Paulin-Henriksson et
al. (2003), as well as a new candidate. We regard the \finals
second-level candidates as less securely established. They need
further analysis over a longer timeline before they can be
convincingly accepted or rejected. We also showed a further \finalt
third-level lightcurves, that lie just below our threshold for
designation as microlensing candidates.

The candidate selection algorithm proceeds by automatically performing
fits to a blended microlensing event and a variable star template for
every lightcurve. Only if the blended microlensing fit is preferred is
the lightcurve retained for further cuts. This is a powerful way of
discarding variable stars, but preserving long-timescale microlensing
events.  However, some short-timescale events with variable baselines
(caused by contamination from nearby variables) are also thrown out.
The selection algorithm also includes explicit cuts on achromaticity
and on the location of events in an analogue of the colour-magnitude
diagram. This is a useful way of discarding long-period variables and
Miras, which given our sampling may fail to repeat during the
three-year baseline. We note that some variable stars are excellent
fits to blended microlensing events in two or three passbands. This is
worrisome for any experiment identifying long-timescale microlensing
events based on single pass-band data.

Our candidate selection algorithm is unrestricted, in the sense that
it has no explicit threshold on timescale or on flux
variation. However, there is an implicit threshold on flux variation
in the last cut, as the signal-to-noise ratio in microlensing must
dominate over the signal-to-noise ratio in the residual
variations. There are some obvious ways of extending the work in this
paper. First, there is a fourth-year of M31 imaging data taken by the
MEGA collaboration, which is now publically available. It can provide
additional discrimination against variable stars for our second-level
candidates.  Second, the efficiency of our survey must be calculated
as a function of threshold to provide physically meaningful
constraints on the baryon fraction of the M31 dark halo.

The great variety of lightcurves in pixel-lensing data implies that
the optimum candidate selection algorithm may have a different
structure, depending on the timescale of the event. For example, for
short-timescale events, the selection algorithm given above could be
changed in the following manner. The datapoints associated with a
candidate bump in the lightcurve are removed and a variable star
template is fitted to the remainder.  The first step in the selection
then proceeds with a comparison between the goodness of fit of a
microlensing event with this bumpy baseline as compared to a variable
star fit to the whole lightcurve. In other words, the two options are
variable star and microlensing contaminated by a nearby variable.  If
the latter is preferred, the lightcurve is retained and subjected to
further analysis.  It may also make sense to fine-tune the details of
the cuts according to each field and CCD. The variable star
contamination varies from CCD to CCD and therefore the optimum
location of the cuts also changes.

The problem of identifying variable objects in pixel lensing is much
more difficult than the corresponding problem in classical
microlensing. Straight line cuts in the projected planes of
multi-dimensional parameter spaces have been employed here and in
other previous searches (e.g., Paulin-Henriksson et al. 2003, de Jong
et al. 2004). However, the decision boundaries separating microlensing
from non-microlensing need not be as simple as hyper-planes in
practice. It would be interesting to set more sophisticated pattern
recognition algorithms, such as neural networks or self-organizing
maps, onto this task (e.g., Belokurov, Evans \& Le Du, 2003, 2004).
In this connection, we note that neural networks to hunt for the
nova-like lightcurves in the POINT-AGAPE dataset have already been
successfully developed by Feeney (2004).

Nonetheless, whatever microlensing candidate selection algorithm is
used, the most important point is that the efficiency of the survey
must be computable. Of course, different selection algorithms may have
different efficiencies, but they must give consistent results for the
characteristic mass and baryon fraction of the microlensing
population.  Here, the need for a well-defined efficiency has caused
us to excise small portions of the data around bright resolved stars,
which give rise to artefacts in the superpixel method. The fraction of
the CCDs so masked is small, but this does mean that some genuine
microlensing events have been excluded.  In a future paper, we will
compute the efficiency of this survey and present constraints on the
dark matter content of the M31 halo.

\begin{table} 
\begin{center}
\begin{tabular}{cccccc}
\hline
Lightcurve & Field/CCD & R & RA & Dec \\
\hline
1 & F1,CCD1  & 12\farcm91 & 00\h43\m52\fs9 & 41\degr16\arcmin59\arcsec \\
2 & F1,CCD1  & 14\farcm19 & 00\h43\m59\fs2 & 41\degr18\arcmin06\arcsec \\
3 & F1,CCD2  &  7\farcm12 & 00\h42\m20\fs8 & 41\degr21\arcmin42\arcsec \\
4 & F1,CCD2  &  5\farcm56 & 00\h42\m30\fs3 & 41\degr21\arcmin01\arcsec \\
5 & F1,CCD4  & 26\farcm14 & 00\h44\m25\fs6 & 41\degr34\arcmin05\arcsec \\
6 & F1,CCD4  & 25\farcm63 & 00\h44\m03\fs6 & 41\degr37\arcmin01\arcsec \\
7 & F1,CCD4  & 24\farcm93 & 00\h44\m02\fs6 & 41\degr36\arcmin17\arcsec \\
8 & F1,CCD4  & 14\farcm86 & 00\h43\m26\fs9 & 41\degr28\arcmin40\arcsec \\
9 & F2,CCD2  & 18\farcm79 & 00\h41\m33\fs6 & 41\degr02\arcmin51\arcsec \\
10 & F2,CCD2 & 23\farcm61 & 00\h41\m20\fs1 & 40\degr58\arcmin39\arcsec \\
11 & F2,CCD3 &  6\farcm06 & 00\h43\m16\fs0 & 41\degr14\arcmin59\arcsec \\
12 & F2,CCD3 &  3\farcm89 & 00\h43\m00\fs4 & 41\degr13\arcmin41\arcsec \\
13 & F2,CCD3 &  4\farcm44 & 00\h42\m48\fs0 & 41\degr11\arcmin44\arcsec \\
14 & F2,CCD4 & 15\farcm60 & 00\h43\m23\fs6 & 41\degr02\arcmin24\arcsec \\
15 & F2,CCD4 & 20\farcm25 & 00\h43\m11\fs3 & 40\degr56\arcmin32\arcsec \\
16 & F2,CCD4 & 22\farcm60 & 00\h42\m44\fs4 & 40\degr53\arcmin32\arcsec \\
\hline
\end{tabular}
\end{center}
\caption{The locations of the \finalt third-level lightcurves in right
ascension and declination (J2000.0), together with the projected
distance $R$ from the centre of M31.}  
\label{tab:locsthree} 
\end{table}

\section*{Acknowledgements}
The Isaac Newton Telescope is operated on the island of La Palma by
the Isaac Newton Group in the Spanish Observatorio del Roque de los
Muchachos of the Instituto de Astrofisica de Canarias.  Work by JA and
YT has been supported through a grant from the Leverhume Trust
Foundation. VB is supported by the Particle Physics and Astronomy
Research Council (PPARC), while MW is supported by a PPARC
studentship.  SCN was supported by the Swiss National Science
Foundation and by the Tomalla Foundation.  AG was supported by grant
AST~02-01266 from the National Science Foundation (US).  CS was
supported by the Indo French Center for Advanced Research (IFCPAR)
under project No. 2404-3.  We would like to thank Maurizio Salaris for
generating the synthetic stellar population datasets used in this
paper.

{}

\appendix

\section{The Photometry Transformation}

\begin{table*}
\begin{center}
\begin{tabular}{lccccccccc}
\hline
Field & CCD & $\Gamma_V$ & $\Gamma_{R_1}$ & $\Gamma_{R_2}$ &
$\Gamma_I$  & $\beta_V$  & $\beta_{R_1}$ & $ \beta_{R_2}$ &
$\beta_I$ \\
\hline
northern & 1  & 24.64 & 24.33 & 24.30 & 23.90 & -0.60 & -0.18 & -0.28
& -0.20\\ 
northern & 2  & 24.63 & 24.32 & 24.27 & 23.88 & -0.61 & -0.21 & -0.27
& -0.15\\
northern & 3 & 24.81 & 24.61 & 24.35 & 23.60  & -0.71 & -0.29 & -0.46
&
-0.23\\
northern & 4 & 24.74 & 24.49 & 24.22 & 23.53  & -0.69 & -0.30 & -0.54
& -0.40\\
southern & 1  & 24.61 & 24.21 & 24.26 & 24.04 & -0.50 & -0.06 & -0.28
& -0.25\\ 
southern & 2  & 24.59 & 24.26 & 24.24 & 23.88 & -0.65 & -0.19 & -0.24
& -0.12\\
southern & 3 & 24.59 & 24.29 & 24.26 & 23.98  & -0.61 & -0.17 & -0.23
&
-0.17\\
southern & 4 & 24.61 & 24.29 & 24.27 & 23.96  & -0.60 & -0.20 & -0.22
& -0.13\\
\hline
\end{tabular}
\end{center}
\caption{The coefficients in the transformations from Sloan to Cousins
magnitudes according to each field and CCD.}
\label{tab:necref}
\end{table*}

In the main text, we frequently convert from Sloan magnitudes to
Cousins, using equations derived by Paulin-Hernriksson (2002).
For the data taken in 1999, the transformations are
\begin{eqnarray}
V &=& \Gamma_V + g + \beta_V (g-r) \nonumber \\
R &=& \Gamma_{R_1} + r + \beta_{R_1} (g-r).
\end{eqnarray}
For the data taken in 2000-2001, the transformations are
\begin{eqnarray}
R &=& \Gamma_{R_2} + r + \beta_{R_2} (r-i) \nonumber \\
I &=& \Gamma_I + i + \beta_I (r-i) 
\end{eqnarray}
The photometric calibration and colour transformation is
worked out independently for each CCD. The coefficients are listed
in Table~\ref{tab:necref}.

\section{The \finalt Third-Level Lightcurves}

The lightcurves of the \finalt third-level events are displayed in
Figure~\ref{fig:vas1}.

\begin{figure*}
\centerline{\epsfxsize=\hsize \epsfbox{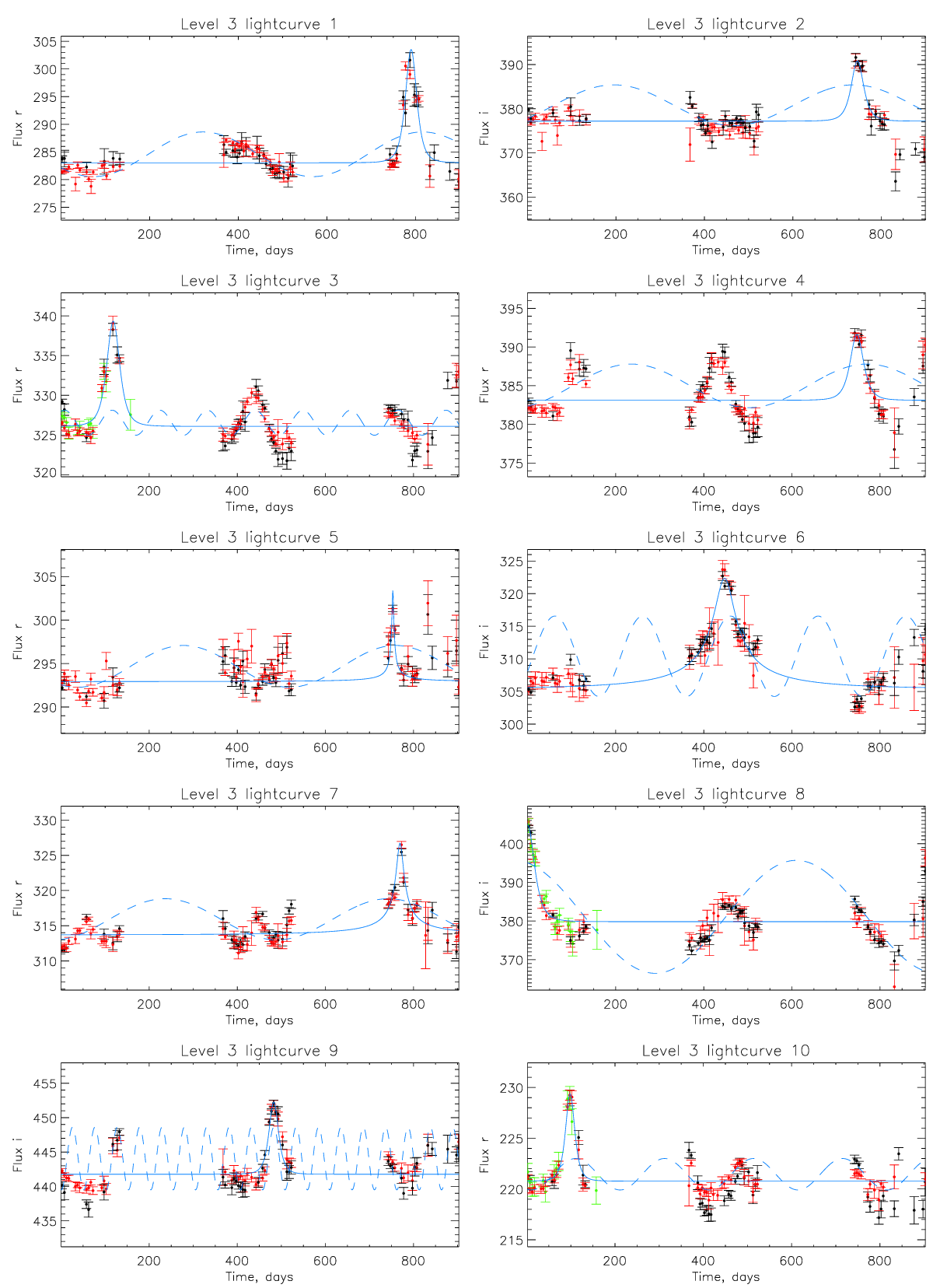}}
\end{figure*}
\begin{figure*}
\centerline{\epsfxsize=\hsize \epsfbox{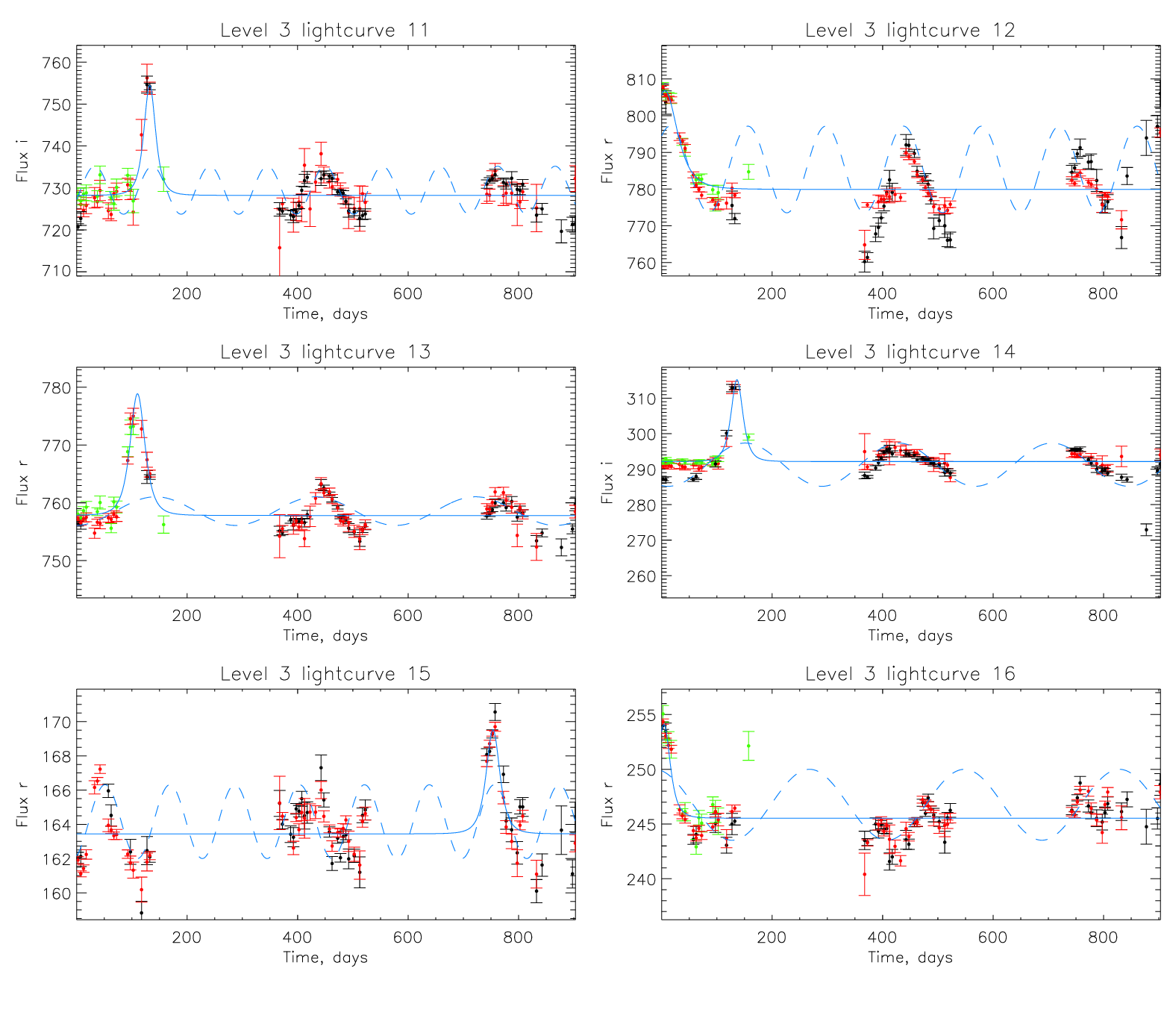}}
\caption{The third-level lightcurves. These just fail our threshold
for designation as microlensing candidates.  Data from all three-bands
are overplotted using the convention: $g$ band as green, $r$ band as
red, and $i$ band as black. Also shown are the variable fit
(blue-dashed) and the global microlensing fit (unbroken blue).}
\label{fig:vas1}
\end{figure*}

\end{document}